\documentclass[manuscript]{aastex}

\shorttitle{CHaMP II. Bolometric Luminosities}
\shortauthors{Ma et al.}

\begin{document}


\title{The Galactic Census of High- and Medium-mass Protostars. \\
II. Luminosities and Evolutionary States of a 
Complete Sample of Dense Gas Clumps}

\author{Bo Ma\altaffilmark{1}, Jonathan C. Tan\altaffilmark{1,2}, Peter J. Barnes\altaffilmark{1}}
\affil{Department of Astronomy, University of Florida, FL, 32611, USA}
\affil{Department of Physics, University of Florida, FL, 32611, USA}

\begin{abstract}
The Census of High- and Medium-mass Protostars (CHaMP) is the first large-scale ($280^\circ<l<300^\circ$, $-4^\circ<b<2^\circ$), unbiased, sub-parsec resolution survey of Galactic molecular clumps and their embedded stars. Barnes et al. (2011) presented the source catalog of $\sim$300 clumps based on HCO$^+$(1-0) emission, used to estimate masses $M$. Here we use archival mid-infrared to mm continuum data to construct spectral energy distributions. Fitting two-temperature grey-body models, we derive bolometric luminosities, $L$. We find the clumps have $10 \lesssim L/L_\odot \lesssim 10^{6.5}$ and $0.1 \lesssim L/M/[L_\odot/M_\odot]\lesssim 10^3$, consistent with a clump population spanning a range of instantaneous star formation efficiencies from 0 to $\sim 50\%$. We thus expect $L/M$ to be a useful, strongly-varying indicator of clump evolution during the star cluster formation process.  We find correlations of the ratio of warm to cold component fluxes and of cold component temperature with $L/M$. We also find a near linear relation between $L/M$ and {\it Spitzer}-IRAC specific intensity (surface brightness), which may thus also be useful as a star formation efficiency indicator. The lower bound of the clump $L/M$ distribution suggests the star formation efficiency per free-fall time is $\epsilon_{\rm ff}<0.2$.  We do not find strong correlations of $L/M$ with mass surface density, velocity dispersion or virial parameter. We find a linear relation between $L$ and $L_{\rm HCO^+(1-0)}$, although with large scatter for any given individual clump. Fitting together with extragalactic systems, the linear relation still holds, extending over 10 orders of magnitude in luminosity. The complete nature of the CHaMP survey over a several kiloparsec-scale region allows us to derive a measurement at an intermediate scale bridging those of individual clumps and whole galaxies.
\end{abstract}



\section{Introduction}

\linespread{1.3}
Stars form from the gravitational collapse of the densest regions of
giant molecular clouds (GMCs). In particular star clusters, likely the
dominant mode of star formation (Lada \& Lada 2003; Gutermuth et
al. 2009), are born from $\sim$ parsec-scale gas {\it clumps} within
GMCs. However, many open questions remain (see, e.g. McKee \& Ostriker
2007; Tan et al. 2012; Hennebelle \& Falgarone 2012). How are GMCs formed out of the diffuse
interstellar medium? Why does star formation occur in only a small
fraction of the available gas in GMCs? What is the star formation rate
and efficiency over the GMC lifetime and what processes control this?
What is the timescale of star cluster formation: is it fast (Elmegreen
2000, 2007) or slow (Tan et al. 2006) with respect to the free-fall
time?  What processes control the evolution and overall star formation
efficiency of a star-forming clump?

To help address some of these open questions, Barnes et al. (2011,
hereafter Paper I) have designed a multi-wavelength survey, the Census
of High- and Medium-mass Protostars (CHaMP). Starting in the 3mm band,
the aim of CHaMP has been to map a complete sample of molecular gas
structures in a $20^\circ \times 6^\circ$ region in the Galactic plane
($280^\circ<l<300^\circ$, $-4^\circ <b < +2^\circ$), and to measure
their associated star formation activity from the near- to
far-IR. Using the 4m Nanten telescope, this region was first surveyed
in the J=1-0 transitions of $\rm ^{12}CO$, $\rm ^{13}CO$, $\rm
C^{18}O$ and $\rm HCO^+$ (Yonekura et al. 2005). This
sequence of species traces progressively higher densities and the
mapping was carried out in this order so as to identify all the
locations of dense gas, without having to map the entire region in the
tracers of the densest gas.
So $^{13}$CO was only observed where the $^{12}$CO integrated
intensity was above $10$~K km s$^{-1}$, and C$^{18}$O and HCO$^+$ were
observed where $^{13}$CO was brighter than $5$~K km s$^{-1}$. 
Then a follow-up campaign was begun to map the dense gas
regions found in the Nanten survey. The follow-up is conducted in a
number of 3mm molecular transitions with the 22m Mopra telescope at
much higher sensitivity and angular resolution than Nanten telescope
(Paper I). This observing strategy distinguishes the CHaMP 
survey from all other Galactic plane surveys of dense gas. 

In Paper I, maps of the CHaMP regions in HCO$^+$(1-0) line
emission observed by the Mopra telescope were presented. A total of
303 massive molecular clumps were identified. This sample has the
following properties: integrated line intensities 1-30 K km~s$^{-1}$,
linewidths 1-9 km s$^{-1}$, FWHM sizes 0.2-2 pc, mean mass surface
densities $\Sigma \sim 0.01$ to $\sim 1$~g cm$^{-2}$ and masses $\rm
\sim 10$ to $\sim 10^4\:M_\sun$.

In this paper we use archival infrared and millimeter data to
investigate the SEDs and luminosities of these HCO$^+$ clumps, with
the goal being to characterize their evolutionary state with respect
to star cluster formation. The paper is organized as follows. \S2
describes the IR and mm data used in this study. \S3 describes our
methods of estimating clump fluxes. \S4 presents our results,
including the clump masses, bolometric fluxes, bolometric
luminosities, luminosity-to-mass ratios, warm and hot component
fluxes, and cold component temperatures and bolometric
temperatures. In particular, we examine the correlation of various
potential tracers of embedded stellar content with the
luminosity-to-mass ratio, and then emphasize the use of this ratio as
an evolutionary indicator for star cluster formation. \S5 presents
further discussion, including searches for potential correlation of
luminosity-to-mass ratio with clump mass surface density and virial
parameter. It also discusses the luminosity versus HCO$^+$ line
luminosity relation from clumps to whole galaxies.  \S6 summarizes our
conclusions.

\section{Infrared and Millimeter Observational Data} 

The first goal of this paper is to measure fluxes at various
wavelengths coming from the CHaMP clumps. Here we describe the main
observational datasets that we use to derive these fluxes.

\subsection{MSX}

The Midcourse Space Experiment (MSX) was launched in April 1996. It
conducted a Galactic plane survey ($0^\circ<l<360^\circ$, $\vert
b\vert < 5^\circ$), which covers all the CHaMP clumps. The four MSX
band wavelengths are centered at 8.28, 12.13, 14.65 and 21.3 \micron.
The best image resolution is $\sim18$\arcsec\ in the $8.28$ $\micron$
band, with positional accuracy of about 2\arcsec. The instrumentation
and survey are described by \citet{egan96}. Calibrated images of the
Galactic plane were obtained from the online MSX image server at the
IPAC website at: http://irsa.ipac.caltech.edu/data/MSX/. For
simplicity, we assume conservative common absolute flux uncertainties
of 20\% for all the IR data ({\it MSX, IRAS, Spitzer IRAC}), similar
to that estimated for {\it IRAS} (M. Cohen, private comm.).


\subsection{IRAS}

The Infrared Astronomical Satellite (IRAS) performed an all sky survey
at 12, 25, 60 and 100 \micron.  The nominal resolution is about
4\arcmin\ at 60\micron.  High Resolution Image Restoration (HIRES)
uses the Maximum Correlation Method \citep[MCM,][]{aumann90} to
produce higher resolution images, better than 1\arcmin\ at 60 \micron.
Sources chosen for processing with HIRES were processed at all four
IRAS bands with 20 iterations. The pixel size was set to
15\arcsec\ with a 1$^\circ$ field centered on the target. The absolute
fluxes of the IRAS data are expected to be accurate to about 20\%.

\subsection{Spitzer IRAC}

The Spitzer InfraRed Array Camera (IRAC) is a four-channel camera that
provides simultaneous $5.2' \times 5.2'$ images at 3.6, 4.5, 5.8, and
8 \micron\ with a pixel size of $1.2'' \times 1.2''$ and angular
resolution of about 2\arcsec\ at 8 \micron. We searched the Spitzer
archive at http://irsa.ipac.caltech.edu/applications/Spitzer/SHA/
for IRAC data near the positions of our HCO$^+$ clumps.
We found IRAC data for 284 out of our 303 clumps. Most of these data
are from two large survey programs: PID 189 (Churchwell, E., ``The
SIRTF Galactic Plane Survey") and PID 40791 (Majewski, S., ``Galactic
Structure and Star Formation in Vela-Carina"). We used the post basic
calibration data to estimate the fluxes of these clumps, which we
assume has a $20\%$ uncertainty.

\subsection{Millimeter data}

Hill et al. (2005) carried out a 1.2-mm continuum emission survey
toward 131 star-forming complexes using the Swedish ESO Submillimetre
Telescope (SEST) IMaging Bolometer Array (SIMBA). SIMBA is a
37-channel hexagonal bolometer array operating at a central frequency
of 250 GHz (1.2mm), with a bandwidth of 50 GHz. It has a half power
beam width of 24\arcsec\ for a single element, and the separation
between elements on the sky is 44 arcsec. Hill et al. list the
1.2-mm flux for 404 sources, 15 of which are in our sample.

\section{Data Analysis}

\subsection{Definition of Clump Angular Area and HCO$^+$ Masses}

Paper I presented maps of the CHaMP region in HCO$^+$(1-0) line
emission using the 22m Mopra telescope, identifying 303 massive
molecular clumps. Elliptical clump sizes were defined based on 2D
Gaussian fitting for each HCO$^+$ clump. The ellipse size quoted in
columns 9 and 10 of Table 4 of Paper I is the FWHM angular size of the
major and minor axes of the Gaussian fit. Clump masses, $M$, were
evaluated based on integrating the derived column density distribution
over the full area of the Gaussian profile ($M_{\rm col}$ listed in
column 9 of their Table 5). Note the derivation of mass surface
densities and masses from the observed HCO$^+$(1-0) intensity depends
on: 
(1) in the view of one of our team (JCT), the conversion of observed 
HCO$^+$(1-0) line intensity to total HCO$^+$ column density is assumed to 
have an uncertainty $\sim 30\%$; in the view of another (PJB), there is 
no identifiable reason for this assumption, since the analysis in Paper 
I showed that there is no such uncertainty, beyond the points mentioned next.
(2) the abundance of HCO$^+$ ($X_{\rm HCO^+}\equiv n_{\rm
  HCO^+}/n_{\rm H2} = 1.0\times 10^{-9}$ was adopted in Paper I, being
a median value from a number of observational and astrochemical
studies). The uncertainty in this mean abundance is itself uncertain:
in this paper we will assume a factor of 2 uncertainty, i.e. a range
of 0.5 to $2.0\times 10^{-9}$ for the mean abundance. In addition,
clump to clump variations in $X_{\rm HCO^+}$ are expected: we will
assume a dispersion of a factor of 2. There may be a number of effects
that lead to systematic variation of $X_{\rm HCO^+}$ with
environmental conditions. For example, we have recently found (Barnes
et al. 2013) that the HCO$^+$ abundance may be enhanced in the
vicinity of ionizing radiation from massive stars, with a possibly
lower $X_{\rm HCO^+}$ in the majority of darker, more quiescent
clumps.  If confirmed, this particular effect would tend to have the
effect of increasing the masses quoted here for the more quiescent
clumps, but decreasing the masses for the minority of vigorously
star-forming clumps. Future work to improve the calibration of
HCO$^+$-derived masses is needed. 
(3) the distance to the sources (the clumps' median distance uncertainty 
is estimated in Paper I to be ~20\% based mostly on classical distance 
estimates to GMC complexes and assuming association of clumps 
with a particular GMC complex. Here we use a slightly larger, more 
conservative value of 30\% for the absolute distance uncertainty 
[see also Paper I for a more extensive discussion of distance estimates], 
i.e. leading to $\sim 60\%$ uncertainties in $M$). 
Combining these uncertainties, we conclude that the absolute
mass estimate of any particular clump may be uncertain by as much as a
factor of $\sim 3$.

To measure the continuum fluxes at various wavelengths coming from the
CHaMP HCO$^+$ clumps, we define the clump size as two times larger
than the FWHM ellipse derived in Paper I, i.e. its radial extent is
equal to 1 FWHM at a given position angle. For a 2D Gaussian flux
distribution as assumed in Paper I, the area inside this ellipse
encloses 93.75\% of the total flux. Thus with this definition of clump
size we expect to enclose close to 100\% of the total HCO$^{+}$ flux
measured in Paper I, and presumably close to 100\% of the continuum
flux associated with each clump.

The clumps are highly clustered in space so that on a scale of two
times the FWHM ellipse the majority of them, $\sim 70$\% of the
sample, suffer from overlap with a neighboring clump ($\sim 30$\%
overlap on the scale of one times the FWHM ellipse). While the
original clump definition from Paper I also used their velocity space
information, sometimes nearby clumps also overlap in velocity to some
extent. We have developed an approximate method to estimate the fluxes
of these clumps where there are image pixels belonging to more than
one ellipse. We first calculate the angular distance, normalized by
the size of the ellipse, from the overlapped pixel to the center of
each clump. This normalized angular distance is defined as $r_{\rm
  norm}= ((d_x/a)^2+(d_y/b)^2)^{1/2}$, where $d_x$ and $d_y$ are the
angular distance from the overlapped pixel to the minor and major axis
of each ellipse, and $a$ and $b$ are the angular sizes of the major
and minor axis of the ellipse. Then the flux of each overlapped pixel
is assigned to its nearest ellipse according to this normalized
angular distance.

\subsection{Clump and Background Flux Measurements}\label{S:method2}

The MSX and IRAS data exist for all 303 CHaMP clumps and these form
the basis for our spectral energy distribution measurements. We
describe here the method we use to derive the fluxes from the clumps
based on these imaging data. We then describe how we utilize the
mid-infrared IRAC data and the mm data where it is available.

Using the coordinates, sizes and geometries of the HCO$^+$ sources,
fluxes were deduced first by directly integrating over the images,
this total flux being expressed as $F_{\nu,{\rm tot}}$. However, we
expect that some fraction of this flux can come from foreground and
background sources along the line of sight that are not associated
with the clump. For simplicity we refer to this foreground and
background emission as the ``background flux'', $F_{\nu,b}$. We
evaluate $F_{\nu,b}$ as the median pixel value in the region between
the clump ellipse (as defined here) and an ellipse twice as large
(i.e. four times the FWHM size of Paper I), excluding areas that are
part of other clumps.

In the end, we derived two fluxes: without and with background
subtracted, which are $F_{\nu,{\rm tot}}$ and $F_\nu = F_{\nu,{\rm
    tot}}-F_{\nu,b}$, respectively.  The error of the fluxes are
estimated from the combination of two terms.  The first is the
uncertainty in the absolute flux from the particular telescope. The
data used here are generally assumed to be accurate to about
$20\%$. The second term is from the background subtraction. Since in
the Galactic plane it is often difficult to estimate the background
emission, we treat the background level as an error term in our flux
error estimation. So the fractional error is $\rm
(0.2^2+(F_{\nu,b}/F_{\nu,{\rm tot}})^2)^{1/2}$. In the following, we
have carried out the analysis for both flux estimates, $F_{\nu,{\rm
    tot}}$ and $F_\nu$.

Next we use a two-temperature grey body model to fit the spectral
energy distribution (SED) in order to estimate the bolometric fluxes,
$F_{\rm tot}$ (no background subtracted) and $F$ (background
subtracted), (calculated by integrating over the fitted SED and
assuming negligible flux escapes in the near-IR and shorter
wavelengths) and temperatures of the clumps, following the method of
\citet{hunter00} and \citet{faundez04}. Each temperature component of
the grey body model is described by:
\begin{equation}
F_\nu = \Omega B_\nu(T)\{ 1-\exp (-\tau_\nu) \}
\end{equation}
where $B_\nu(T)=(2h\nu^3/c^2)/[\exp(h\nu/kT)-1]$ is the Planck function 
for the black body flux density (where $c$ is the speed of light, $h$ is the Planck constant 
and $k$ is the Boltzmann constant), $\Omega$ is the angular size of the source, 
and $T$ is the temperature. The dependence of the optical depth, 
$\tau_\nu$, with frequency, $\nu$, is given by:
\begin{equation}
\tau_\nu = \left(\frac{\nu}{\nu_0}\right)^\beta ,
\end{equation} 
where $\beta$ is the emissivity index and $\nu_0$ is the turnover frequency.

In this fitting procedure, for the colder component (subscript ``c''),
we explored parameter values in the ranges $T_c=10-50$~K and
$\beta_c=1.0-2.5$. These values of $\beta_c$ are those expected from
laboratory experiments and observational results (see Schnee et
al. 2010 and references therein). Also, \citet{faundez04} found
$\beta_c$ to be in this range for their sample of sources. We find
$\nu_0$ to generally be in the range $3-30$~THz. For the warmer
component (subscript ``w''), $T_w$ was allowed to have values in the
range $100-300$~K, while $\beta_w$ was fixed to 1 following
\citet{hunter00} and \citet{faundez04}. The choice of $\beta_w=1$ is
motivated both by theoretical calculations and by observational
evidence \citep[][pp. 201-203]{whittet92}. The angular size of the
colder component, $\Omega_c$ was set equal to the angular size of the
clump, including accounting for reduction due to overlap with other
clumps.  For the warmer component the angular size, $\Omega_w$, is
derived from the best fitting result, always being smaller than the
angular size of the clump.

\begin{figure}
 \includegraphics[width=17cm]{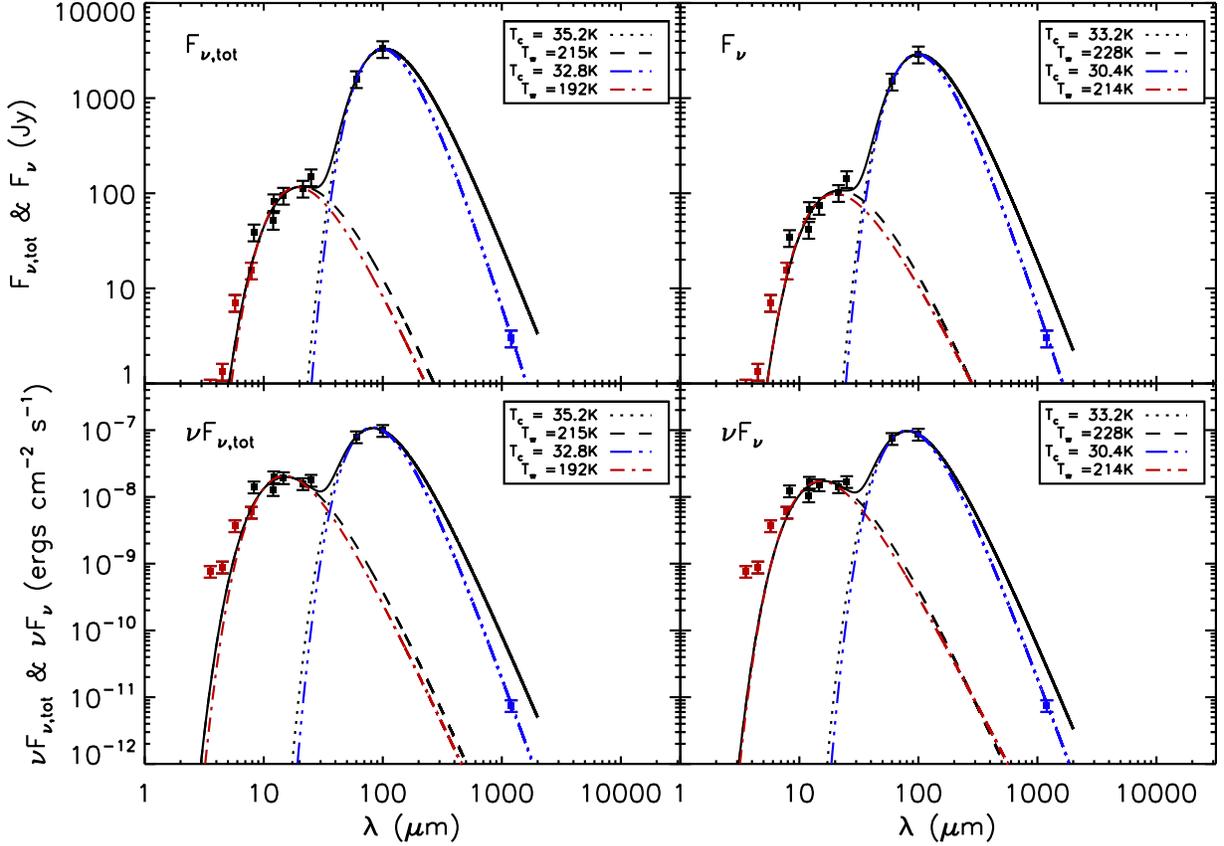}
\vspace{-1.4cm}
\caption{\footnotesize
SED fitting results of BYF 73 (G286.2+0.2). {\it (a) Top left:} $F_{\rm \nu,tot}$
(no background subtracted); {\it (b) Top right:} $F_{\rm
  \nu}$ (background subtracted);  {\it (c) Bottom left:} $\nu F_{\rm
  \nu,tot}$ (no background subtracted); {\it (d) Bottom right:} $\nu F_{\rm
  \nu}$ (background subtracted). The data in order of increasing
wavelength are Spitzer-IRAC, MSX, IRAS, SIMBA. In the fiducial case,
which is used for the main analysis of the 303 CHaMP clumps, we only
use MSX and IRAS data to find colder component temperatures $T_c=35.2,
33.2$~K (without and with background subtraction) (dotted lines) and
$T_w=215, 228$~K (dashed lines). The totals are shown by the solid
lines. Fitting MSX, IRAS and the mm SIMBA flux leads to revised model
fits with $T_c=32.8, 30.4$~K (dash-dot-dot-dotted lines). Fitting IRAC, MSX
and IRAS leads to revised model fits with $T_w=192, 214$~K (dash-dotted
lines). In both cases the bolometric fluxes change by $\lesssim 5\%$
from the fiducial case.
\label{fig:byf73}}
\end{figure}

The values of $T_c$ are not particularly well constrained by the IRAS
data, which extend to a longest wavelength of only $100\:{\rm \mu
  m}$. For those 15 sources where we do have mm fluxes reported from
SEST-SIMBA, we examine how the two-temperature grey body model fit
changes when we do make use of the mm flux. Note, for the mm fluxes,
not having access to estimates of $F_{\nu,b}$, we assume that
background subtraction makes a negligible difference,
i.e. $F_{\nu,b}\ll F_\nu$.  In Fig.~\ref{fig:byf73} we present the SED
and model fits of BYF 73 (G286.2+0.2), which is one of the more massive and
actively star-forming clumps in the sample (Barnes et al. 2010), as
one example to show the effect of the mm flux measurement. The results
from only the MSX and IRAS data are: $T_c=35.2,33.2$~K,
$\beta_c=1.45,1.52$ and $\nu_0=101,31.1$~THz, without and with
background subtraction, respectively. Adding in the mm flux we now
derive $T_c = 32.8, 30.4$~K, $\beta_c = 1.82, 1.78$ and
$\nu_0=15.4,11.1$~THz for these same cases. The bolometric flux,
obtained by integrating over the model spectrum, changes from $(1.32,
1.22) \times 10^{-7}$ erg$^{-1}$ s$^{-1}$ cm$^{-2}$, without and with
background subtraction respectively, to $(1.30, 1.20) \times10^{-7}$
erg$^{-1}$ s$^{-1}$ cm$^{-2}$ when the mm flux is utilized.

The fitting results for all 15 sources with mm flux measurements using
the no background subtraction method are summarized in
Table~\ref{tab:mm}.  These results show that $T_c$ typically changes
by $\lesssim 5$~K after including the mm flux.\footnote{We note
  that all 15 of these sources contain, or lie near,
  luminous young clusters.  Therefore it is possible that the
  remainder of the clumps, many of which are relatively quiescent in
  their star-forming activity, may have systematically lower Tc when
  good longer-wavelength data are included, than this subsample would
  suggest.} The mean value changes
by about $10\%$. We find $\beta_c$ changes from $2.0\pm0.33$ to
$1.85\pm0.41$ after utilizing the mm flux. Most importantly, we find
the bolometric fluxes, $F_{\rm tot}$ and $F$, typically change by
$\lesssim 10\%$ after including the mm flux. Thus we conclude that the
lack of longer wavelength data for the main sample only introduces a
modest uncertainty of $\sim 10\%$ in $F_{\rm tot}$ and $F$. Note,
however, that the limited FIR/sub-mm coverage of the SEDs, even with
the SEST-SIMBA data, prevent accurate measurement of $T_c$, $\nu_0$
and $\beta$. This situation will be improved with forthcoming data
from the {\it Herschel} Hi-GAL survey (Molinari et al. 2010).

\begin{deluxetable}{ccccc}
\tablewidth{0pc}
\tablecaption{Effect of 1.2~mm data on SED fitting.  \label{tab:mm}}
\tablehead{
\colhead{BYF} & \colhead{$T_{c}$\tablenotemark{a}}  & \colhead{$\log(F)$\tablenotemark{a}} & \colhead{$T_{c}$\tablenotemark{b}} & {$\log(F)$\tablenotemark{b}} \\
 \colhead{No.}& \colhead{($\rm K$)} & \colhead{(erg s$^{-1}$ cm$^{-2}$)} &  \colhead{($\rm K$)} &  \colhead{(erg s$^{-1}$ cm$^{-2}$)}}
\startdata
 73 & 35.2 &     $-6.88$ & 32.8 & $-6.89$ \\
126a & 37.9 & $-5.64$ &37.9 & $-5.64$ \\
 126b & 35.2 & $-6.33$ &37.2 & $-6.33$ \\
 126c & 30.4 & $-6.39$ & 31.8 & $-6.45$ \\
 128a & 36.4 & $-7.10$ &34.5 & $-7.04$ \\
 128b & 39.0 & $-5.80$ &39.9 & $-5.79$ \\
 131b & 37.0 & $-7.07$ &39.5 & $-7.06$ \\
 131c & 50.0 & $-6.46$ &41.4 & $-6.50$ \\
 131d & 39.6 & $-6.13$ &41.2 & $-6.13$ \\
 131e & 49.2 & $-6.32$ &42.7 & $-6.33$ \\
 132d & 37.0 & $-5.92$ &39.8 & $-5.89$ \\
 132e & 31.0 & $-6.68$ &37.5 & $-6.61$ \\
 162 & 31.1  &  $-8.46$ &33.5 & $-8.52$ \\
 163a & 40.1 & $-7.12$ &39.4 & $-7.12$ \\
 163b & 35.5 & $-7.48$ &39.2 & $-7.49$ \\
\enddata
\tablenotetext{a}{using MSX and IRAS data}
\tablenotetext{b}{using MSX, IRAS \& mm data}
\end{deluxetable}

In this paper, we choose not to use the IRAC data for our fiducial SED
fitting (although we do examine certain correlations of clump
properties with the flux in the IRAC bands). The IRAC data are not
available for about 10\% of the CHaMP clumps (generally those furthest
from the midplane) and we wish to maintain the same procedure for all
the clumps in the sample. Furthermore, the bolometric luminosity is
dominated by the colder component, even for clumps with the most
active star formation (see, e.g. Fig.~\ref{fig:byf73}). For BYF~73,
when we compare SED fitting (no background subtraction) with just
MSX+IRAS to that with IRAC+MSX+IRAS we see: $T_c$ changes from
$35.1$~K to $35.0$~K, $T_w$ changes from $215$~K to $230$~K and $F$
changes from $1.323\times10^{-7}$ erg$^{-1}$ s$^{-1}$ cm$^{-2}$ to
$1.328\times10^{-7}$ erg$^{-1}$ s$^{-1}$ cm$^{-2}$.

\section{Results} 

\subsection{HCO$^+$ Masses}

\begin{figure}
\includegraphics[width=17cm]{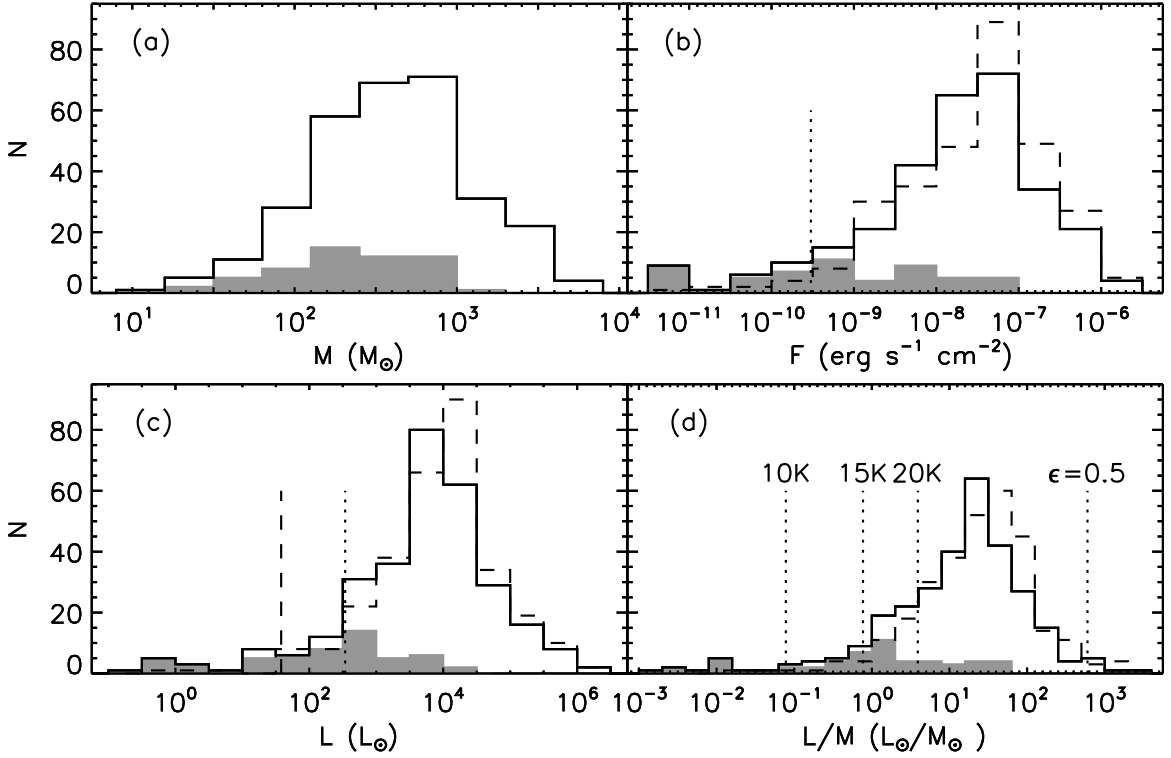}
\vspace{-1.7cm}
\caption{
\footnotesize {\it (a) Top left:} Distribution of the masses ($M$,
estimated from HCO$^+$(1-0)) of the 303 CHaMP clumps. The grey shaded
histogram shows the sources for which the bolometric flux, $F$,
measurements are uncertain due to background subtraction (see
(b)). {\it (b) Top right:} Distribution of the bolometric fluxes ($F$,
{\it solid line}, estimated from the 2 temperature greybody fit to the
background subtracted SED; $F_{\rm tot}$, {\it dashed line}, estimated
from the 2 temperature greybody fit to the total SED [no background
  subtracted]). The grey shaded histogram shows the sources for which the
bolometric flux, $F$, measurements are uncertain due to having IRAS
100\micron\ background fluxes $ >0.75$ of the clump flux). The vertical
dotted line shows a bolometric flux of $3\times10^{-10}$ erg$^{-1}$
s$^{-1}$ cm$^{-2}$, which is our estimate for the 10$\sigma$
sensitivity flux limit of the IRAS data for a clump with typical
angular size of 60\arcsec. {\it (c) Bottom left:} Distribution of
bolometric luminosities ($L$, {\it solid line}, estimated from $F$;
$L_{\rm tot}$, {\it dashed line}, estimated from $F_{\rm tot}$). The grey
shaded histogram, a subset of $L$, shows the same sources as described
in (b) with uncertain flux measurements due to background
subtraction. The vertical dashed and dotted lines show the luminosity
corresponding to the flux limit shown in (b) for clumps at 2.0 and 6.0
kpc, respectively. {\it (d) Bottom right:} Distribution of luminosity
to mass ratios ($L/M$, {\it solid line}; $L_{\rm tot}/M$, {\it dashed
  line}). The grey shaded histogram, a subset of $L/M$, shows the same
sources as described in (b) with uncertain flux measurements due to
background subtraction. Three vertical dotted lines on the left side show 
$L/M=0.078, 0.77, 3.9$~$L_\odot/M_\odot$ (from left to right),
which corresponds to a grey-body with $T=10, 15, 20$~K. The 
vertical dotted line on the right side shows $L/M=600$~$L_\odot/M_\odot$, 
which corresponds a clump with an equal mass of gas and stars (i.e., a 
star formation efficiency $\epsilon \equiv M_*/(M_*+M) = 0.5$) that are on the ZAMS.
}
\label{Ldist}
\end{figure}

In this paper we set the clump mass, $M$, equal to that derived from
analysis of HCO$^+$(1-0) emission, $M_{\rm col}$ (listed in column 
9 of Table 5, Paper I). The distribution of these masses is presented 
in Fig.~\ref{Ldist}a.
The masses range from $\sim 10 - 10^4\:M_\odot$, with mean of
$723\:M_\odot$ and median of $427\:M_\odot$. The clump masses and
other clump properties are also listed in Table~\ref{tab:clumps}. 
Additional, secondary clump properties are listed in Table~\ref{tab:clumps2}.

As discussed above, uncertainties in absolute clump mass are likely to
be at the level of about a factor of 4, mainly due to uncertainties in
HCO$^+$ abundance. We expect relative clump masses are somewhat better
determined, especially since a large fraction of the CHaMP clumps are
in the Carina sprial arm, with about half in the same $\eta$ Carinae
giant molecular association at a common distance of $\sim 2.5$~kpc.

\subsection{Bolometric Fluxes}

The bolometric flux distributions without, $F_{\rm tot}$ and with,
$F$, background subtraction are presented in Fig.~\ref{Ldist}b. The
mean 10$\sigma$ sensitivity of the 4 IRAS bands are 0.7, 0.65,
0.85 and 3.0 Jy, which correspond to a bolometric flux of about
$3\times10^{-10}$ erg$^{-1}$ s$^{-1}$ cm$^{-2}$ for a source with a
typical angular size of 60\arcsec. This limit is also shown in
Fig.~\ref{Ldist}b. We see that $F_{\rm tot}$ can be detected at better
than $10\sigma$ for nearly all of the CHaMP clumps. We assume the
uncertainty in $F_{\rm tot}$ is about 20\% from the absolute flux
calibration of the IR observations and about 10\% from the two
temperature greybody model fitting, i.e. adding in quadrature to about
22\%.

For the faintest clumps, the total flux from the direction of the
clump, $F_{\rm tot}$, can be similar to that of the background
(i.e. the region surrounding the clump). The background subtracted
flux, $F_\nu$, can thus be very small (or even formally negative) at a
particular wavelength. The uncertainty assigned to $F_\nu$ is of order
the same level as the background. For deriving bolometric fluxes, the
flux at 100\micron\ is typically most important. Thus we flag those
clumps that have a 100\micron\ background flux that is $>0.75$ times
the clump flux, and consider these values of $F$, $L$ and $L/M$ to be
highly uncertain, i.e. $\gtrsim 100\%$ uncertainties.

\subsection{Bolometric Luminosities}

Given the clump distances from Paper I and our derived bolometric
fluxes, we calculate the bolometric luminosities $L_{\rm tot}$ and $L$
(without and with background subtraction, respectively). The
distributions of $L_{\rm tot}$ and $L$ are shown in Fig.~\ref{Ldist}c.

Adopting a typical distance uncertainty of 30\% as explained in \S~3.1, we
then estimate an uncertainty in $L_{\rm tot}$ of about 64\%. $L$ has
somewhat greater uncertainty due to background flux estimation, and
again we flag those sources where we expect this source of error
dominates.

The mean luminosities are $ \langle L_{\rm tot} \rangle =5.2\times
10^4 L_\odot$ and $ \langle L \rangle =4.2\times10^4 L_\odot$. For
reference, this is about the luminosity of a 20~$M_\odot$ zero age
main sequence (ZAMS) star (Schaller et al. 1992). The median values of
$L_{\rm tot}$ and $L$ are $1.06\times 10^4 L_\odot$ and $6.2\times 10^3
L_\odot$, respectively: half of the sample are lower in luminosity
than a single 12~$M_\odot$ ZAMS star.

Note that the previous surveys of dust emission toward massive star
forming regions by \citet{mueller02} and \citet{faundez04} found
$ \langle L_{\rm tot} \rangle = 2.5\times10^5 L_\odot$ and $2.3\times 10^5 L_\odot$,
respectively. These values are much larger that those of the CHaMP
clumps. We attribute this difference as being due to the different
selection criteria of the samples: CHaMP is a complete sample of dense
gas independent of star formation activity, while these other surveys
were selected based on (massive) star formation indicators.



\subsection{Luminosity-to-mass ratios}\label{S:LoverM}

During the evolution of star-forming clumps, i.e. the formation of
star clusters, the gas mass will decrease due to incorporation into
stars and dispersal by feedback, causing the luminosity-to-mass ratio
to increase. So $L/M$ should be an evolutionary indicator of the star
cluster formation process. The distribution of $L/M$ is shown in
Fig.~\ref{Ldist}d.

Three dotted vertical lines at $L/M=0.078, 0.77,
3.9$~$L_\odot/M_\odot$ are used to show the values expected of clouds
with dust temperatures of $ T =10, 15, 20$~K, which can be achieved in
starless clumps via external heating, as evidenced by temperature
measurements of Infrared Dark Clouds (e.g. Pillai et al. 2006). These
values are calculated via
\begin{equation}
\label{eq:LoverM_OH94}
\frac{L/M}{L_\odot/M_\odot} = \frac{4\pi}{\Sigma} \int B_\nu (1-{\rm exp}(-\tau_\nu) \:{\rm d}\nu \rightarrow 0.0778 \left(\frac{ T }{10\:{\rm K}}\right)^{5.65},
\end{equation}
where the latter evaluation is based on integrating the opacities of
the Ossenkopf \& Henning (1994) moderately-coagulated thin ice mantle
dust model (and adopting a gas-to-dust mass ratio of 155) for clouds
with $0.01<\Sigma / {\rm g\:cm^{-2}} < 1$ and $10< T /{\rm K}<20$
(there is a modest dependence of $L/M$ on $\Sigma^{0.02}$, which we
ignore, normalizing the numerical factor of eq.~(\ref{eq:LoverM_OH94})
to $\Sigma=0.03\:{\rm g\:cm^{-2}}$, typical of the CHaMP clump
sample). Values of $L/M \sim 1 L_\odot/M_\odot$ are thus expected to
define the lower end of the $L/M$ distribution, as is observed.



To understand the upper end of the observed distribution, consider a
clump with an equal mass of gas and stars that are on the zero age
main sequence (ZAMS). For a Salpeter IMF down to 0.1$M_\odot$, this
will have $L/M \sim 600 L_\odot / M_\odot$ \citep{leitherer99, tan02}. 
Other IMFs typically considered for Galactic
star-forming regions give similar numbers to within about a factor of
two. This value is close to the upper end of the distribution of $L/M$
shown in Fig.~\ref{Ldist}d. Note that as the gas mass goes to very
small values, $L/M$ should rise far above $600 L_\odot /
M_\odot$. However, in this case a smaller fraction of the bolometric
luminosity will be re-radiated in the MIR and FIR, and so would be
missed by our analysis. Also such ``revealed'' clusters with small
amounts of dense gas would not tend to be objects in the CHaMP sample, which
is complete only on the basis of emission of dense gas tracers.


\begin{figure}
\includegraphics[width=16cm]{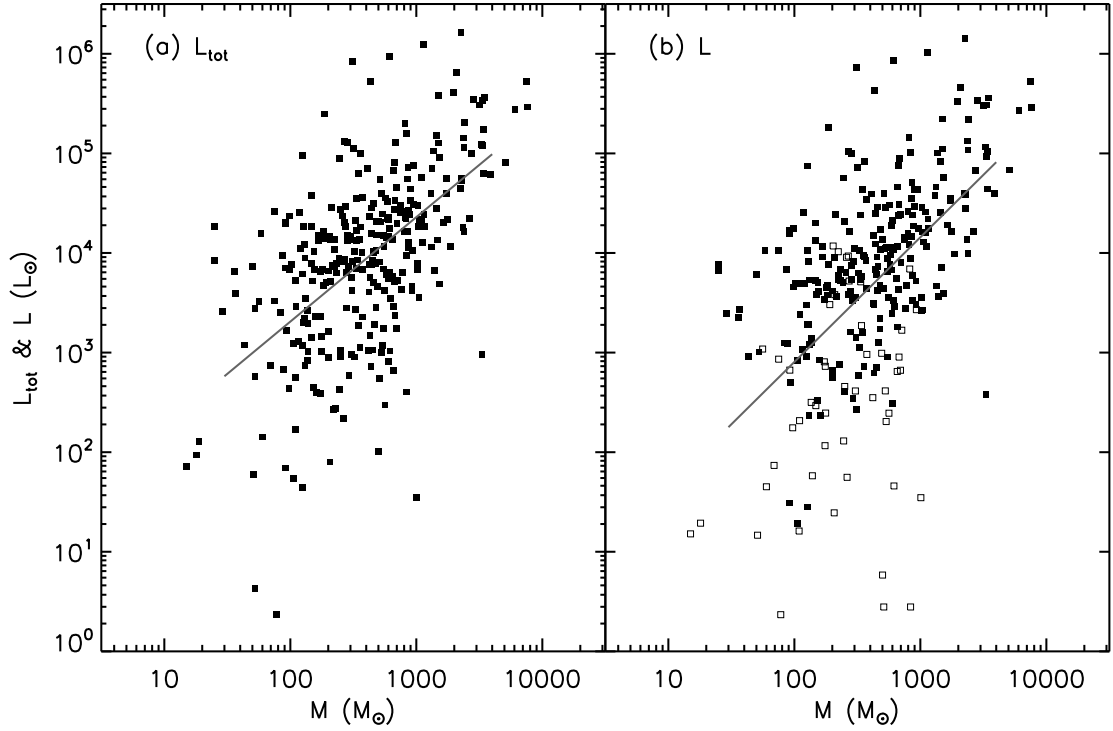}
\vspace{-1.0cm}
\caption{
\footnotesize {
{\it (a) Left:} Correlation of $L_{\rm tot}$ with $M$, with best-fit
relation $L_{\rm tot}/L_\odot = 16.2\times (M/M_\odot)^{1.05}$ shown
with a Spearman rank correlation coefficient 0.54 and
$1.2\times10^{-24}$ probability for a chance correlation.  {\it (b)
  Right:} Correlation of $L$ with $M$, with best-fit relation
$L/L_\odot = 3.0\times (M/M_\odot)^{1.25}$ shown with a Spearman rank
correlation coefficient 0.55 and $7.2\times10^{-25}$ probability for a
chance correlation. Open squares show clumps with uncertain
measurements of $F$ due to IRAS 100\micron\ background subtraction
(see Fig.~\ref{Ldist}b). Note, these are still used to help define the
correlation; their larger uncertainties lead to an asymmetric
distribution of points about the best fit relation.
}}
\label{fig:lm}
\end{figure}

To investigate the relation between bolometric luminosity and gas mass
(i.e. how luminosity depends on mass), we also show the correlation
between $L_{\rm tot}$ and $M$ in Fig.~\ref{fig:lm}a and the
correlation between $L$ and $M$ in Fig.~\ref{fig:lm}b.  The best-fit
power law results (e.g., following methodology of Kelly 2007) are as
follows:
\begin{equation}
L_{\rm tot}/L_\odot = 16.2(\pm9.5)\times (M/M_\odot)^{1.05\pm0.09}
\end{equation}
with Spearman rank correlation coefficient $r_s=0.54$ and probability
for a chance correlation $p_s \ll 10^{-4}$ (formally
$p_s=1.2\times10^{-24}$, but this value depends sensitively on the
assumed shape of the tails of the distribution functions, which are
not well-defined for real datasets) for the no background subtraction
method and
\begin{equation}
L/L_\odot = 3.0(\pm1.8)\times (M/M_\odot)^{1.25\pm0.11}
\end{equation}
with $r_s=0.55$ and $p_s \ll 10^{-4}$ (formally
$p_s=7.2\times10^{-25}$; note the open symbols in Fig.~\ref{fig:lm}b
have larger uncertainties, explaining the asymmetric distribution of
points about the best fit relation) for the background subtraction
method.  Both show significant positive correlations. The more massive
the clump is, the more luminous it tends to be.

The mean, median and standard deviation of $\log (L_{\rm tot}/M /
[L_\odot/M_\odot])$ are 1.34, 1.43 and 0.77 respectively for
non-background subtraction method.  For the background subtraction
method, the mean, median and standard deviation of $\log(L/M/
[L_\odot/M_\odot])$ are 1.06, 1.25 and 0.97 respectively.

\citet{molinari08} have studied the SEDs of 42 potentially massive
individual young stellar objects (YSOs). By fitting the SEDs with YSOs
models they obtained the bolometric luminosity and envelope mass,
$M_{\rm env}$. They presented the $L_{\rm bol}-M_{\rm env}$ diagram as
a tool to diagnose the pre-MS evolution of massive YSOs. For their
sample, the mean, median and standard deviation of $\log(L/M)$ are
1.91, 1.77 and 0.66 respectively.  This illustrates the different
nature of their sample: objects that are already forming massive stars
and with much higher values of $L/M$. However, we caution that
systematic differences could also arise because of the different
methods being used to derive masses (i.e. HCO$^+$ versus mm flux-based masses).


Similarly, \citet{mueller02}, \citet{beuther02} and \citet{faundez04}
reported mean values of $\log(L/M)$ as $2.04\pm0.34$, $1.18\pm0.34$
and $1.75\pm0.38$.  Note here that in \citet{beuther02} they have used
opacity from \citet{hildebrand83}, which is
4.9 times smaller than 
the opacity from \citet{ossen94} used in \citet{mueller02} and 
\citet{faundez04}. 
So the mass derived in \citet{beuther02} would be 4.9 times 
smaller and their mean $\log(L/M)$ will be $1.87\pm0.34$ 
if they adopt the opacity from \citet{ossen94}.


\subsection{The Warm Component}

\begin{figure}
\includegraphics[width=20cm]{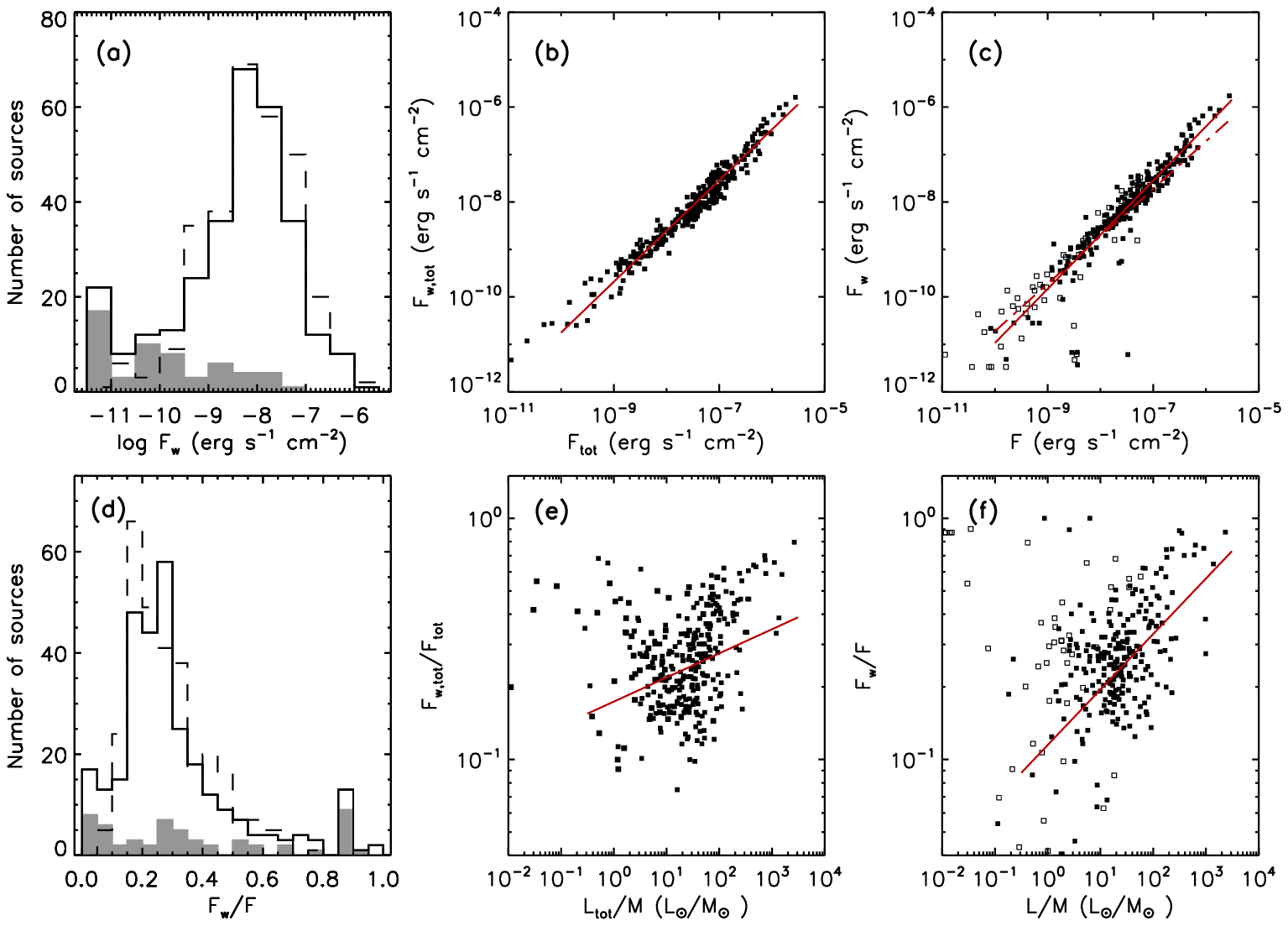}
\vspace{-1.5cm}
\caption{\footnotesize 
{\it (a) Top left:} Distributions of $F_w$ (solid line) and $F_{w,{\rm
    tot}}$ (dashed line). The shaded histogram shows the sources for
which the bolometric flux, $F$, measurements are uncertain due to
background subtraction (see Fig. \ref{Ldist}b). {\it (b) Top middle:}
Correlation of $F_{w,{\rm tot}}$ with $F_{\rm tot}$, with a best-fit
relation of $F_{w,{\rm tot}} = 0.89\times F_{\rm tot}^{1.08}$ with
$r_s=0.98$ and a negligible value of $p_s$.  {\it (c) Top right:}
Correlation of $F_w$ with $F$, with two best-fit relations shown
$F_{\rm w} = 2.69(\pm1.25)\times F^{1.14\pm0.02}$ (solid line) and
$F_{\rm w} = 0.30(\pm0.01)\times F$ (dot-dashed line). Open squares show
clumps with uncertain measurements of $F$ due to IRAS
100\micron\ background subtraction (see Fig.~\ref{Ldist}b). {\it (d)
  Bottom left:} Distribution of $F_w/F$ (solid line) and $F_{w,{\rm
    tot}}/F_{\rm tot}$ (dashed line), with shaded sources as in
(a). {\it (e) Bottom middle:} Correlation of $F_{w,{\rm tot}}/F_{\rm
  tot}$ with $L_{\rm tot}/M$, with best-fit relation $F_{w,{\rm
    tot}}/F_{\rm tot} = 0.19\times (L_{\rm tot}/M/
[L_\odot/M_\odot])^{0.10}$ shown with $r_s=0.29$ and $p_s\ll10^{-4}$
(formally $p_s=4.3\times10^{-7}$). {\it (f) Bottom right:} Correlation
of $F_w/F$ with $L/M$, with best-fit relation $F_{\rm w}/F=
0.11\times(L/M/ [L_\odot/M_\odot])^{0.23}$ shown with $r_s=0.43$ and
$p_s\ll 10^{-4}$ (formally $p_s=2.5\times10^{-12}$). As star cluster
formation proceeds to higher values of $L/M$, the warmer component
becomes more important.}
\label{fig:Fw}
\end{figure}

From the two temperature fitting process, we derived the total,
$F_{\rm w, tot}$, and background-subtracted, $F_{\rm w}$, flux for the
warm component. The distributions of $F_{\rm w,tot}$ and $F_{\rm w}$
are shown in Fig.~\ref{fig:Fw}a. The correlation of $F_{\rm w,tot}$
with $F_{\rm tot}$ is shown in Fig.~\ref{fig:Fw}b, and that of $F_{\rm
  w}$ with $F$ in Fig.~\ref{fig:Fw}c. These both show significant
correlations. We derive a best-fit power law fit for the dependence of
$F_{\rm w, tot}$ on $F_{\rm tot}$,
finding
\begin{equation}
F_{\rm w, tot} = 0.89(\pm0.35)\times F_{\rm tot}^{1.08\pm0.02}.
\end{equation}
For the background subtracted case, which we consider to be the most
accurate measure of the intrinsic properties of the clumps, we try two different constrained fits, finding:
\begin{eqnarray}
F_{\rm w} & = & 2.69(\pm1.25)\times F^{1.14\pm0.02}    \\
F_{\rm w} & = & 0.30(\pm0.01)\times F  
\end{eqnarray}




The distributions of $F_{\rm w,tot}/F_{\rm tot}$ and $F_{\rm w}/F$ are
shown in Fig.~\ref{fig:Fw}d. The warm component flux generally
accounts for $10\%-30\%$ of the total flux, so $F_w$ and $F$ are not
independent, which can contribute to these correlations.

To investigate if there are any systematic trends associated with the
warm component during star cluster formation as measured by the clump
luminosity to mass ratio, we show the correlation of $F_{\rm w,
  tot}/F_{\rm tot}$ versus $L_{\rm tot}/M$ in Fig.~\ref{fig:Fw}e and
$F_w/F$ versus $L/M$ in Fig.~\ref{fig:Fw}f. 

The power law fit results of this positive correlation are as follows:
\begin{equation}
F_{\rm w, tot}/F_{\rm tot} = 0.19(\pm0.03)\times (L_{\rm tot}/M/ [L_\odot/M_\odot])^{0.10\pm0.03}
\end{equation}
and
\begin{equation}
F_{\rm w}/F=0.11(\pm0.02)\times(L/M/ [L_\odot/M_\odot])^{0.23\pm0.02}
\end{equation}
The Spearman rank correlation coefficients (see Fig.~\ref{fig:Fw})
indicates a positive correlation exists in both cases.

Our findings support the idea that as stars gradually form in
molecular clumps and the luminosity-to-mass ratio increases, a larger
fraction of the bolometric flux will emerge at shorter
wavelengths. The specific functional form of this correlation is a constraint on
radiative transfer models of star cluster formation. 


\subsection{The Hot (IRAC Band) Component}

\begin{figure}
\includegraphics[width=20cm]{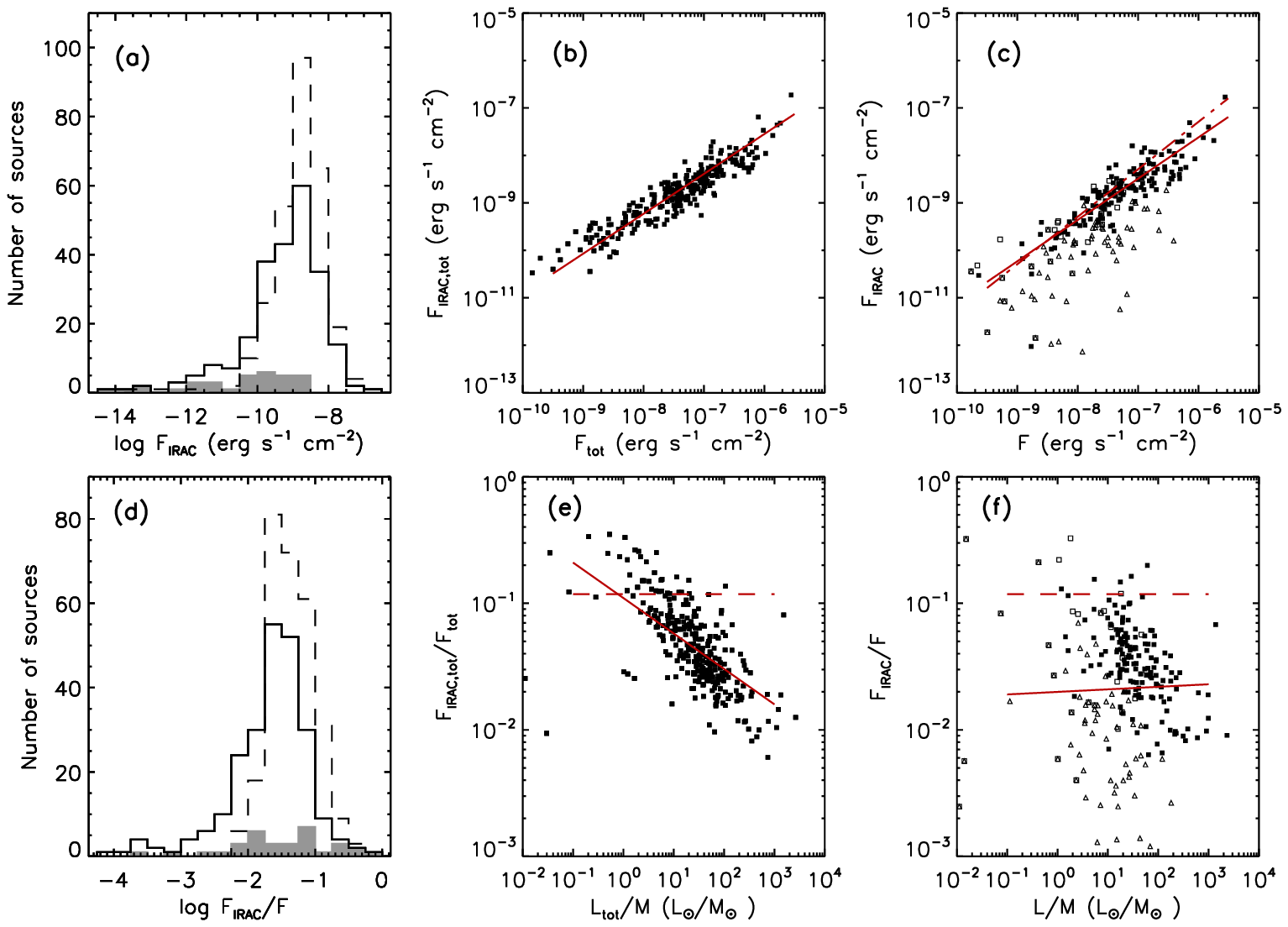}
\vspace{-1.5cm}
\caption{\footnotesize 
{\it (a) Top left:} Distributions of $F_{\rm IRAC}$ (solid line) and
$F_{\rm IRAC, tot}$ (dashed line). The shaded histogram shows the
sources for which the bolometric flux, $F$, measurements are uncertain
due to background subtraction (see Fig. \ref{Ldist}b). {\it (b) Top
  middle:} Correlation of $F_{\rm IRAC, tot}$ with $F_{\rm tot}$, with
a best-fit relation of $F_{\rm IRAC, tot} = 3.1\times10^{-3} F_{\rm
  tot}^{0.84}$. Here $r_s=0.91$ and $p_s$ is negligible.  {\it (c) Top
  right:} Correlation of $F_{\rm IRAC}$ with $F$, with two best-fit
relations shown: $F_{\rm IRAC} = 4.0(\pm3.0)\times10^{-3}\times
F^{0.87\pm0.04}$ (solid line) and $F_{\rm IRAC} =
5.1(\pm0.6)\times10^{-2}\times F$ (dot-dashed line).  Open squares show
clumps with uncertain measurements of $F$ due to IRAS
100\micron\ background subtraction (see \ref{Ldist}b). Open triangles
show clumps with uncertain measurements of $F_{\rm IRAC}$ due to IRAC
8\micron\ background subtraction.
{\it (d) Bottom left:} Distribution of $F_{\rm IRAC}/F$ (solid line)
and $F_{\rm IRAC, tot}/F_{\rm tot}$ (dashed line), with shaded sources
as in (a). {\it (e) Bottom middle:} Correlation of $F_{\rm IRAC,
  tot}/F_{\rm tot}$ with $L_{\rm tot}/M$, with a best-fit relation of 
$F_{\rm IRAC, tot}/F_{\rm tot} =0.11(\pm0.01)\times(L_{\rm tot}/M/ [L_\odot/M_\odot])^{-0.28\pm0.02}$.
Here $r_s=-0.69$ and $p_s$ is negligible. 
The horizontal dashed line corresponds to 0.11, which is the $F_{\rm
  IRAC}/F$ ratio of the dust emission from the diffuse interstellar
medium and is calculated using the data from Li \& Draine (2001).
{\it (f) Bottom right:} Correlation of $F_{\rm IRAC}/F$ with $L/M$,
with a best-fit relation of $F_{\rm IRAC, tot}/F_{\rm tot}
=0.02(\pm0.002)\times(L/M/ [L_\odot/M_\odot])^{0.02\pm0.05}$.  Here
$r_s= -0.14$ and $p_s=0.05$, so there is no significant dependence of
$F_{\rm IRAC, tot}/F_{\rm tot}$ with $L/M$.}
\label{fig:FIRAC}
\end{figure}

We now search for any correlation of the IRAC band flux, which extends
from $\sim 3-9\:{\rm \mu m}$, with the bolometric flux and the
luminosity to mass. These relatively short wavelengths are more
sensitive to hot dust directly heated by embedded young stars. We
first measure the total IRAC band flux using a simple trapezoidal rule
integration in the four IRAC bands, without background subtraction,
$F_{\rm IRAC, tot}$, and then subtract the background to
derive $F_{\rm IRAC}$.


The distributions of $F_{\rm IRAC, tot}$ and $F_{\rm IRAC}$ are shown
in Fig.~\ref{fig:FIRAC}a. The correlation of $F_{\rm IRAC,tot}$
with $F_{\rm tot}$ is shown in Fig.~\ref{fig:FIRAC}b, and that of $F_{\rm
  IRAC}$ with $F$ in Fig.~\ref{fig:FIRAC}c. These both show highly significant
correlations. The power law fit results of these two correlations are as follows:
\begin{equation}
F_{\rm IRAC, tot} = 3.1(\pm1.7)\times10^{-3}\times F_{\rm tot}^{0.84\pm0.03}
\end{equation}
and, trying two constrained fits,
\begin{eqnarray}
F_{\rm IRAC} & = & 4.0(\pm3.0)\times10^{-3}\times F^{0.87\pm0.04}  \\
F_{\rm IRAC} & = & 5.1(\pm0.6)\times10^{-2}\times F.
\end{eqnarray}


The distributions of $F_{\rm IRAC,tot}/F_{\rm tot}$ and $F_{\rm
  IRAC}/F$ are shown in Fig.~\ref{fig:FIRAC}d. The IRAC component flux
generally accounts for $\sim 1\%-10\%$ of the total flux, so $F_{\rm
  IRAC}$ and $F$ are essentially independent, unlike for $F_w$ (above).




To investigate if there are any systematic trends associated with the
IRAC (hot) component during star cluster formation as measured by the clump
luminosity to mass ratio, we show the correlation of $F_{\rm IRAC,
  tot}/F_{\rm tot}$ versus $L_{\rm tot}/M$ in Fig.~\ref{fig:FIRAC}e and
$F_{\rm IRAC}/F$ versus $L/M$ in Fig.~\ref{fig:FIRAC}f. 
The best-fit power law relations are as follows:
\begin{equation}
F_{\rm IRAC, tot}/F_{\rm tot} =0.11(\pm0.01)\times(L_{\rm tot}/M/ [L_\odot/M_\odot])^{-0.28\pm0.02}.
\end{equation}
The Spearman rank correlation coefficient of $F_{\rm IRAC, tot}/F_{\rm
  tot}$ versus $L_{\rm tot}/M$ is negative. We expect this is due to
the fact that $F_{\rm tot}$ and $L_{\rm tot}$ are correlated, while
$F_{\rm IRAC, tot}$ is often dominated by ``background'' (i.e. both
background and foreground, i.e. unrelated) emission.

Attempting a power law fit for $F_{\rm IRAC}/F$ versus $L/M$, we find
\begin{equation}
F_{\rm IRAC}/F =0.02(\pm0.002)\times(L/M/ [L_\odot/M_\odot])^{0.02\pm0.05},
\end{equation}
but with $r_s=-0.14$ and $p_s=0.05$, indicating there is not
significant correlation. So there is no evidence for an increase in
the relative importance of the hot component as cluster evolution (as
measured by $L/M$) proceeds. As the luminosity input into the clump
rises, a fairly constant fraction emerges in the IRAC bands. Again,
this result can provide a constraint on theoretical models of star
cluster formation.

\begin{figure}
\includegraphics[width=16cm]{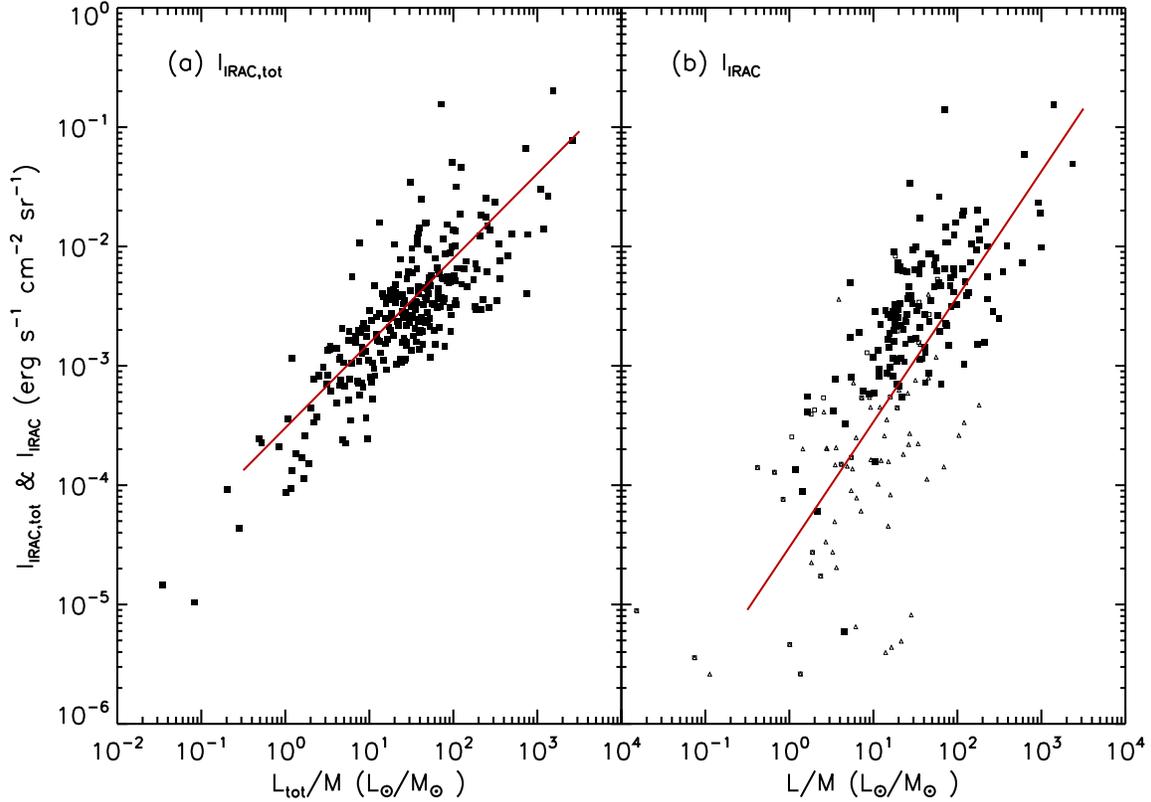}
 \caption{\footnotesize
{\it (a) Left:} Correlation of $I_{\rm IRAC, tot}$ with $L_{\rm
  tot}/M$, with a best-fit relation of 
  $I_{\rm IRAC, tot} = 3.0(\pm0.5)\times10^{-4} \times (L_{\rm tot}/M / [L_\odot/M_\odot])^{ 0.71(\pm0.05)}$ 
 with $r_s=0.57$ and negligible value of $p_s$.  {\it (b) Right:}
 Correlation of $I_{\rm IRAC}$ with $L/M$, with a best-fit relation of
 $I_{\rm IRAC} = 3.0(\pm0.7)\times10^{-5} \times (L/M /
 [L_\odot/M_\odot])^{ 1.05(\pm0.05)}$ with $r_s=0.66$ and negligible
 value of $p_s$. Open squares show clumps with uncertain measurements
 of $F$ due to IRAS 100\micron\ background subtraction (see
 \ref{Ldist}b). Open triangles show clumps with uncertain measurements
 of $F_{\rm IRAC}$ due to IRAC 8\micron\ background subtraction.
}
\label{fig:I_IRAC}
\end{figure}

In order to more directly probe the evolution of IRAC-traced hot dust
emission and its possible correlation with luminosity to mass ratio,
we also calculated the IRAC band specific intensity (surface
brightness) without, $I_{\rm IRAC, tot}$ and with, $I_{\rm IRAC}$
background subtraction (Fig.~\ref{fig:I_IRAC}).
Note that both the specific intensities and the luminosity to mass
ratios are essentially independent of distance uncertainties.
The best-fit relations are as follows:
\begin{equation}
I_{\rm IRAC, tot} = 3.0(\pm0.5)\times10^{-4} \times (L_{\rm tot}/M / [L_\odot/M_\odot])^{ 0.71(\pm0.05)} \:{\rm \rm erg\;s^{-1}\;cm^{-2}\;sr^{-1}} 
\end{equation}
with $r_s=0.57$ and $p_s \ll 10^{-4}$ (formally $p_s=10^{-13}$), and,
trying two constrained fits,
\begin{eqnarray}
I_{\rm IRAC} & = & 3.0(\pm0.7)\times10^{-5} \times (L/M / [L_\odot/M_\odot])^{ 1.05(\pm0.05)} \:{\rm \rm erg\;s^{-1}\;cm^{-2}\;sr^{-1}} \\
I_{\rm IRAC} & = & 5.0(\pm1.9)\times10^{-5} \times (L/M / [L_\odot/M_\odot]) \:{\rm \rm erg\;s^{-1}\;cm^{-2}\;sr^{-1}}. \label{eq:IIRAC_lin}
\end{eqnarray}
The former has $r_s=0.66$ and $p_s\ll 10^{-4}$ (formally
$p_s=10^{-19}$).

Thus the IRAC band specific intensity, which is essentially
independent of $L/M$ (since only a very small fraction of $L$ emerges
at these wavelengths) and more directly traces embedded stellar
populations, has a significant correlation with $L/M$, thus validating 
the use of $L/M$ as an evolutionary indicator of star cluster formation.
The specific functional form of the correlation is a constraint on
radiative transfer models of star cluster formation. 

The near linear relation of $I_{\rm IRAC}$ with $L/M$ (although with
large scatter, which may be expected from IMF sampling) suggests that
$I_{\rm IRAC}$ also has a near linear dependence on embedded stellar
content relative to gas mass, i.e. the instantaneous star formation
efficiency, $\epsilon^\prime \equiv M_*/M$, which, note, is normalized
by the gas mass. (We define $\epsilon \equiv M_*/(M_*+M)$, which
becomes similar to $\epsilon^\prime$ when $\epsilon^\prime \ll 1$.)
Thus, for a Salpeter IMF down to 0.1$M_\odot$ (see \S\ref{S:LoverM}),
\begin{equation}
\epsilon^\prime \simeq 1.0 \frac{L/M}{600 L_\odot/M_\odot} \simeq 0.33(\pm0.16) \frac{I_{\rm IRAC}}{10^{-2}\:{\rm erg\;s^{-1}\;cm^{-2}\;sr^{-1}}},
\label{eq:epsilonIRAC}
\end{equation}
where we have used the numerical result of the constrained linear fit
(eq.~\ref{eq:IIRAC_lin}). This may be a useful relation for estimating
star formation efficiencies of statistical samples of star-forming
clumps (at least those with similar densities to local Galactic
clumps), when only IRAC data are available and a background
subtraction can be performed.

\subsection{Cold Component Dust Temperature and Bolometric Temperature}

\begin{figure}
\includegraphics[width=21cm]{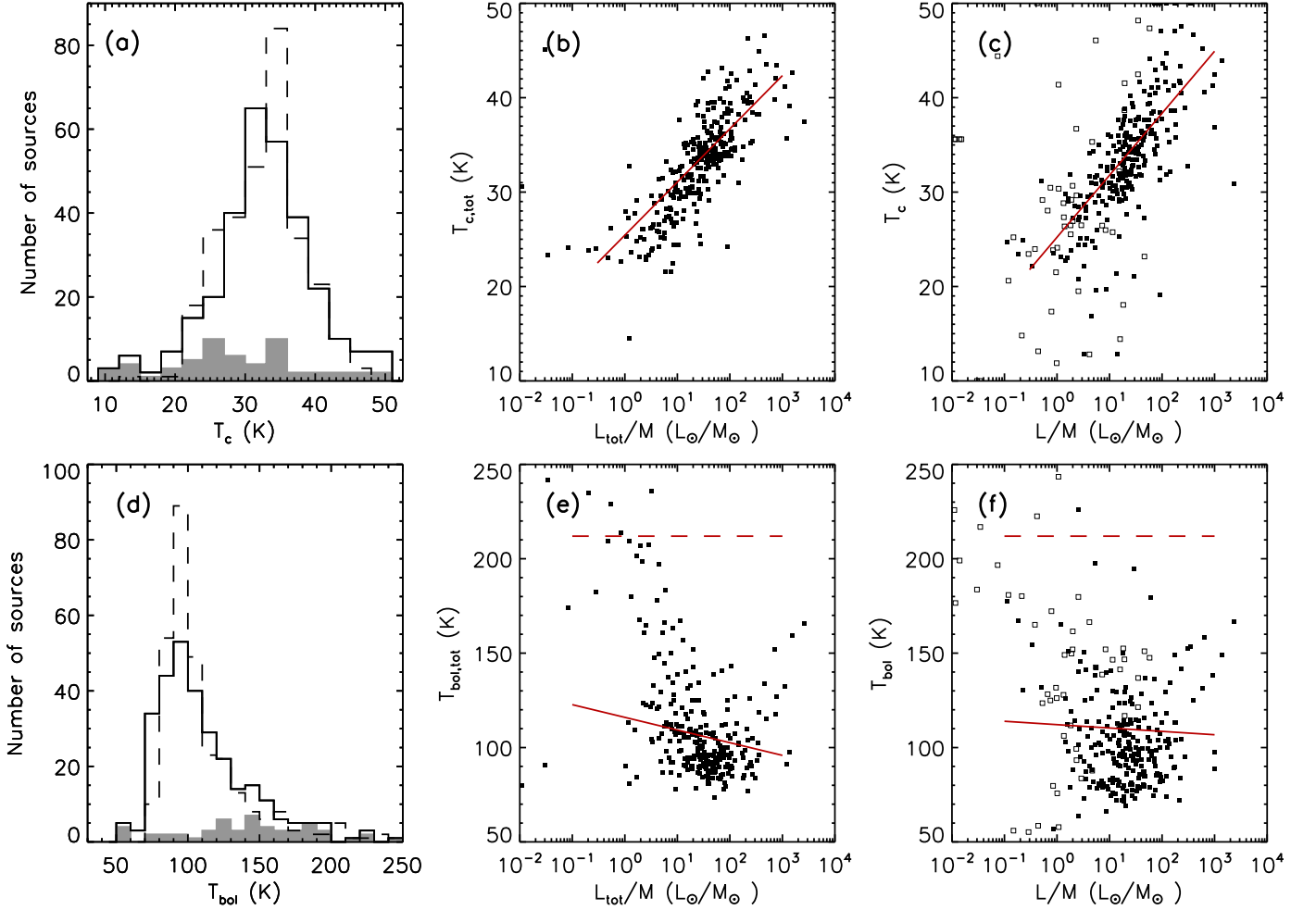}
\vspace{-1.5cm}
\caption{\footnotesize 
{\it (a) Top left:} Distribution of $T_c$ (solid line) and $T_{c,{\rm
    tot}}$ (dashed line). The shaded histogram shows the sources for
which the bolometric flux, $F$, measurements are uncertain due to
background subtraction (see Fig. \ref{Ldist}). {\it (b) Top middle:}
Correlation of $T_{c,{\rm tot}}$ with $L_{\rm tot}/M$, with a best-fit
relation of $T_{c,{\rm tot}}/{\rm K} = 5.6(\pm0.5) \times \log (L_{\rm
  tot}/M / [L_\odot/M_\odot])+25.4(\pm0.8)$, $r_s=0.81$ and negligible
value of $p_s$.  {\it (c) Top right:} Correlation of $T_{c}$ with
$L/M$, with a best-fit relation of $T_c/{\rm K} = 6.6(\pm0.6)\times
\log (L/M / [L_\odot/M_\odot])+25.2(\pm1.0)$ with $r_s=0.65$ and
$p_s=4.8\times10^{-38}$.  Open squares show clumps with uncertain
measurements of $F$ due to IRAS 100\micron\ background subtraction
(see \ref{Ldist}b).  {\it (d) Bottom left:} Distribution of $T_{\rm
  bol}$ (solid line) and $T_{\rm bol,tot}$ (dashed line). The shaded
histogram shows the sources for which the bolometric flux, $F$,
measurements are uncertain due to background subtraction (see
Fig. \ref{Ldist}).  {\it (e) Bottom middle:} $T_{\rm bol,tot}$ versus
$L_{\rm tot}/M$, which does not show a significant correlation (the
best-fit relation of $T_{\rm bol,tot}/{\rm K} =-6.7(\pm3.4) \times
\log (L_{\rm tot}/M / [L_\odot/M_\odot])+116.0(\pm6.4)$ has
$r_s=-0.15$ and $p_s=0.06$). The horizontal dashed line represents
$T=210$ K, which is the bolometric temperature of the dust emission in
the diffuse ISM calculated using the data from Li \& Draine (2001).
{\it (f) Bottom right:} $T_{\rm bol}$ versus $L/M$, which also does
not show a significant correlation (the best-fit relation of $T_{\rm
  bol}/{\rm K} =-1.8(\pm3.3) \times \log (L/M /
[L_\odot/M_\odot])+112.1(\pm3.3)$ has $r_s=-0.15$ and $p_s=0.06$.)  }
\label{fig:Tc}
\end{figure}

We now search for any dependence of the cold component dust
temperature, $T_{\rm c, tot}$ (based on total fluxes with no
background subtracted) and $T_c$ (based on fluxes after background
subtraction), with the luminosity to mass ratio. We note that the
available data for the clumps generally are limited at long
wavelengths to the IRAS 100~$\rm \mu m$ data and so our accuracy for
estimating $T_c$ is limited to about $\pm 5$~K (see
\S\ref{S:method2}).

The distributions of $T_{\rm c, tot}$ and $T_c$ are shown in
Fig.~\ref{fig:Tc}a. The mean values are $ \langle T_{\rm c,tot} \rangle  =33\pm 5$ K
and $ \langle T_{\rm c} \rangle  =33\pm 7$~K. 
These results are similar to those derived in other surveys, such as:
$ \langle T \rangle  = 29\pm9$~K in the large sample of \citet{mueller02};
$ \langle T \rangle  = 45 \pm 11$~K in the large sample of \citet{sri02};
$ \langle T \rangle = 32\pm5$~K in the sample of \citet{molinari00};
$ \langle T \rangle = 35 \pm 6$ K in the sample of \citet{hunter00};
$ \langle T \rangle  = 30$K in the sample of \citet{molinari08};
and $ \langle T \rangle  = 32$K in the sample of F\'aundez et al. (2004).



The correlation of $T_{c,{\rm tot}}$ with $L_{\rm tot}/M$ is shown in
Fig.~\ref{fig:Tc}b and that of $T_{\rm c}$ with $L/M$ in
Fig.~\ref{fig:Tc}c. We see clear positive correlations are present ---
the temperature rises as $L/M$ increases.  We find best-fit relations:
\begin{equation}
T_{c,{\rm tot}}/{\rm K} =  5.6(\pm0.5) \times \log (L_{\rm tot}/M / [L_\odot/M_\odot])+25.4(\pm0.8)
\end{equation}
with $r_s=0.81$ and negligible value of $p_s$, and
\begin{equation}
T_c/{\rm K} =  6.6(\pm0.6)\times \log (L/M / [L_\odot/M_\odot])+25.2(\pm1.0)
\end{equation}
with $r_s=0.65$ and negligible value of $p_s$.

``Bolometric temperature'', $T_{\rm bol}$, has been proposed as a
measure of the evolutionary development of a young stellar object
(YSO) (Ladd et al. 1991; Myers \& Ladd 1993; Myers et al. 1998). It is
the temperature of a blackbody having the same weighted mean frequency
as the observed SED.  As the envelopes in YSO systems are dispersed,
their bolometric temperatures will rise. This is because the FIR
emission decreases while the NIR and MIR emission increases.

We calculated the bolometric temperature for our molecular clumps
following \citet{myers93}:
\begin{equation}
T_{\rm bol} = 1.25\times 10^{-11}  \langle\nu\rangle \;\;  {\rm K \, Hz}^{-1}  \label{Tbol} 
\end{equation}
where $\langle\nu\rangle \equiv \int_0 ^\infty \nu F_\nu d\nu / \int_0
^\infty F_\nu d\nu$ is the flux weighted mean frequency. The
coefficient of $\langle\nu\rangle$ in eq.~(\ref{Tbol}) is chosen so
that a blackbody emitter at temperature $T$ has $T_{\rm bol} = T$.

The distributions of $T_{\rm bol,tot}$ (based on total fluxes with no
background subtraction) and $T_{\rm bol}$ (based on fluxes after
background subtraction) are shown in Fig.~\ref{fig:Tc}d. These have
mean values $92\pm18$~K and $113\pm44$~K, respectively.
For comparison, \citet{mueller02} find a mean value of $78\pm21$ K
for their sample.

The correlation of $T_{\rm bol, tot}$ with $L_{\rm tot}/M$ is shown in
Fig.~\ref{fig:Tc}e and that of $T_{\rm bol}$ with $L/M$ in
Fig.~\ref{fig:Tc}f. We do not find significant correlations, since the
best-fit relations are
\begin{equation}
T_{\rm bol,tot}/{\rm K} =-6.7(\pm3.4) \times \log (L_{\rm tot}/M / [L_\odot/M_\odot])+116.0(\pm6.4)
\end{equation}
with $r_s=-0.15$ and $p_s=0.06$, and
\begin{equation}
T_{\rm bol}/{\rm K} =-1.8(\pm3.3) \times \log (L/M / [L_\odot/M_\odot])+112.1(\pm3.3)
\end{equation}
with $r_s=-0.15$ and $p_s=0.06$.
We suspect the lack of significant correlation is because the
uncertainties in deriving $T_{\rm bol}$ are relatively large compared
to the expected size of any trend for $T_{\rm bol}$ to increase during
star cluster formation. This is in contrast to the measures $F_w/F$
and $I_{\rm IRAC}$, which show clear changes by about a factor of 10
or more as $L/M$ increases.

\section{Discussion}

\subsection{Dependence of $L$ and $L/M$ on Mass Surface Density, $\Sigma$}

\begin{figure}
\includegraphics[width=15cm,height=16cm]{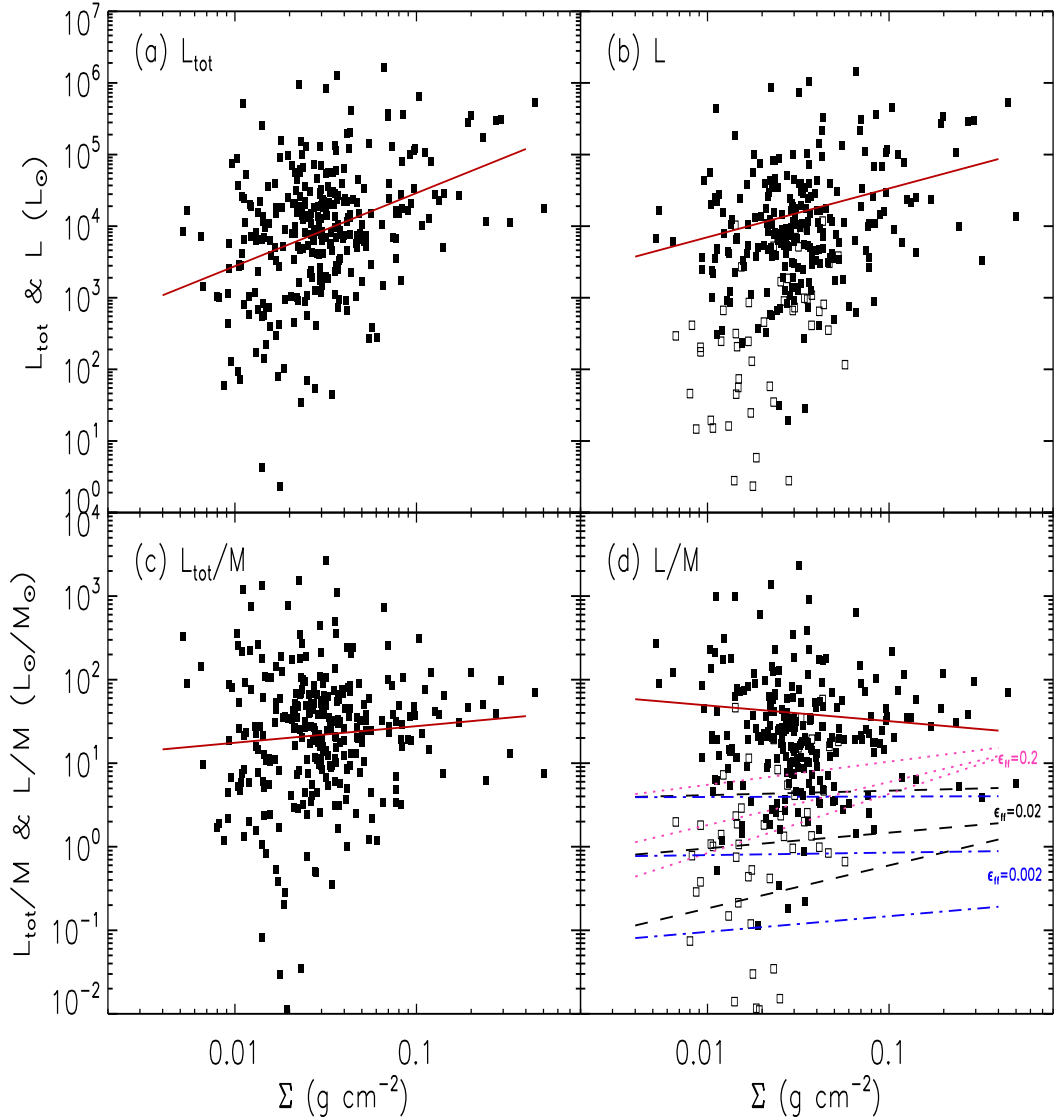}
  \caption{\footnotesize 
  {\it (a) Top Left:} Correlation of $L_{\rm tot}$ with clump mass surface density $\Sigma$. Solid line shows the best-fit relation (see text).
{\it (b) Top Right:} Correlation of $L$ with $\Sigma$. Solid line shows the best-fit relation (see text). Open squares show clumps with uncertain measurements 
 of $F$ due to IRAS 100\micron\ background subtraction (see Fig.~\ref{Ldist}b).
{\it (c) Bottom Left:} Correlation of $L_{\rm tot}/M$ with $\Sigma$. Solid line shows the best-fit relation (see text). 
{\it (d) Bottom Right:} Correlation of $L/M$ with $\Sigma$.
Solid line shows the best-fit relation (see text). The three (black) dashed lines show 
the minimum $L_{\rm min}/M$ expected from only ambient heating and accretion 
luminosity for clumps $M=10^3\:M_\odot$, $ T =10,15,20$~K (from bottom to top)
forming stars at fixed $\epsilon_{\rm ff}=0.02$ (eq.~\ref{eq:LoverM}). 
The three (blue) dash-dotted lines show the minimum $L_{\rm min}/M$ with 
mass $10^3M_\odot$, $ T =10,15,20$~K (from bottom to top) 
forming stars at $\epsilon_{\rm ff}=0.002$.
The three (magenta) dotted lines show the minimum $L_{\rm min}/M$ with 
mass $10^3M_\odot$, $ T =10,15,20$~K (from bottom to top) 
forming stars at $\epsilon_{\rm ff}=0.2$.
Open squares show clumps with uncertain measurements 
 of $F$ due to IRAS 100\micron\ background subtraction (see Fig.~\ref{Ldist}b). 
}
\label{fig:Surfacedensity}
\end{figure}

Consider a clump that forms stars at a fixed efficiency per free-fall
time, $\epsilon_{\rm ff}$. The overall accretion rate to stars is
\begin{equation}
\label{eq:mdot}
\dot{M}_* = \epsilon_{\rm ff} \frac{M}{t_{\rm ff}} = \frac{(8G)^{1/2}}{\pi^{1/4}} \epsilon_{\rm ff} (M\Sigma)^{3/4} = 2.92\times 10^{-4} \frac{\epsilon_{\rm ff}}{0.02} \left(\frac{M}{10^3\:M_\odot} \frac{\Sigma}{\rm g\:cm^{-2}}\right)^{3/4} M_\odot\:{\rm yr^{-1}},
\end{equation}
where we have normalized to a value of $\epsilon_{\rm ff}$ estimated by Krumholz \& Tan (2007). Then the accretion luminosity is
\begin{equation}
\label{eq:Lacc}
L_{\rm acc} = f_{\rm acc} \frac{G\dot{M}_* \bar{m}_* }{\bar{r}_*} = 2270 f_{\rm acc} \frac{\bar{m}_*}{M_\odot} \frac{4R_\odot}{\bar{r}_*} \frac{\epsilon_{\rm ff}}{0.02} \left(\frac{M}{10^3\:M_\odot} \frac{\Sigma}{\rm g\:cm^{-2}}\right)^{3/4} L_\odot.
\end{equation}
Here $f_{\rm acc}$ is the fraction of the accretion power that is
radiated. While for individual protostars we expect $f_{\rm acc}\sim
0.5$ because of the mechanical luminosity of protostellar outflows, in
early-stage star-forming clumps much of the outflow kinetic energy is
likely to be liberated via radiative shocks and thus contribute to the
total clump luminosity. Thus we adopt $f_{\rm acc}=1$ as a fiducial
value. In the above equation, $\bar{m}_*$ is the mean protostellar
mass, weighted by accretion energy release. For a Salpeter IMF from
0.1 to 120~$M_\odot$, the mean stellar mass is 0.353~$M_\odot$, while
the mean gravitational energy is $2.06GM_\odot^2/\bar{r}_*$, assuming
$\bar{r}_*$ is independent of $m_*$ (discussed below). For accretion
near the end of individual star formation, this implies
$\bar{m}_*\simeq 1.4M_\odot$, however the typical unit of accretion
energy release will be when the protostar has $2^{-1/2}$ of its final
mass. Thus we estimate $\bar{m}_*\simeq 1 M_\odot$ as a typical
fiducial value in eq.~(\ref{eq:Lacc}). 

The protostellar evolution models of \citet{tan02}, developed
for protostars forming with accretion rates appropriate for cores
fragmenting from a clump with $\Sigma\simeq 1\:{\rm g\:cm^{-2}}$ (see
also Stahler 1988; Palla \& Stahler 1992; Nakano et al. 2000; McKee \&
Tan 2003), indicate that the sizes of all protostars are close to
$\sim 3$ to $4 R_\odot$ when their masses are $\lesssim
1M_\odot$. After this the size increases along the deuterium core
burning sequence, reaching about $6\:R_\odot$ by the time the
protostars have $1.5\:M_\odot$. After this, sizes stay relatively
constant until $m_*\sim 5\:M_\odot$. Given these relatively modest
changes in $r_*$ with $m_*$, we adopt a fiducial value of
$\bar{r}_*=4\:R_\odot$ in eq.~(\ref{eq:Lacc}).

We can now use eqs.~(\ref{eq:Lacc}) and (\ref{eq:LoverM_OH94}) to estimate minimum values of $L/M$
for star-forming clumps. 
We have
\begin{eqnarray}
\frac{L_{\rm min}/M}{L_\odot/M_\odot} & = & 0.77 \left(\frac{ T }{15\:{\rm K}}\right)^{5.65} + \frac{L_{\rm acc}/M}{L_\odot/M_\odot}\\
 & = & 0.77 \left(\frac{ T }{15\:{\rm K}}\right)^{5.65} + 2.27 f_{\rm acc} \frac{\bar{m}_*}{M_\odot} \frac{4R_\odot}{\bar{r}_*} \frac{\epsilon_{\rm ff}}{0.02} \left(\frac{M}{10^3\:M_\odot}\right)^{-1/4} \left(\frac{\Sigma}{\rm g\:cm^{-2}}\right)^{3/4},
\label{eq:LoverM}
\end{eqnarray}
where $ T $ is the dust temperature expected from ambient heating of
starless clumps. Note that because of internal stellar luminosities that will 
contribute in addition to $L_{\rm acc}$, $L_{\rm min}/M$ provides only 
a lower bound on the distribution of $L/M$ of star-forming clumps.

In Fig.~\ref{fig:Surfacedensity}a and b, we plot the dependence of
$L_{\rm tot}$ and $L$ with $\Sigma$. Note $\Sigma$, like $M$, is
based on the HCO$^+$ observations and analysis. We estimate
$\Sigma$ as $M/2$ divided by the projected area of the FWHM ellipse of
Paper I.
This will give a value of $\Sigma$ for the typical mass
element in the clump. We find best-fit relations:
\begin{equation}
L_{\rm tot} = 3.15(\pm1.33)\times10^5\times (\Sigma/{\rm g\:cm^{-2}})^{1.03\pm0.15} \:L_\odot
\end{equation}
with $r_s=0.33$ and $p_s \ll 10^{-4}$ (formally $p_s=6\times10^{-9}$), and
\begin{equation}
L =  1.70(\pm0.76) \times 10^5\times(\Sigma/{\rm g\:cm^{-2}})^{0.70\pm0.18} \:L_\odot
\end{equation}
with $r_s=0.34$ and $p_s\ll 10^{-4}$ (formally $p_s=2\times10^{-9}$).

In Fig.~\ref{fig:Surfacedensity}c and d, we plot the dependence of
$L_{\rm tot}/M$ and $L/M$ with $\Sigma$. We do not find any evidence
for a correlation, since the best-fit relations are
\begin{equation}
L_{\rm tot}/M = 44.1(\pm17.0) \times (\Sigma/{\rm g\:cm^{-2}})^{0.20\pm0.14}\:L_\odot/M_\odot
\end{equation}
with $r_s=0.07$ and $p_s=0.26$, and
\begin{equation}
L/M = 20.4(\pm5.5) \times (\Sigma/{\rm g\:cm^{-2}})^{-0.19\pm0.14}\:L_\odot/M_\odot
\end{equation}
with $r_s=-0.12$ and $p_s=0.14$. 

One caveat of the above results is that $L/M$ and $\Sigma$ are
inversely correlated via $M$, and this may be making it more difficult
to discern any rise of $L/M$ with $\Sigma$. We note that high $\Sigma$
clumps, e.g. with $\Sigma>0.1\:{\rm g\:cm^{-2}}$ all have $L/M\gtrsim
4 L_\odot/M_\odot$. We also considered our other ``good'' cluster
evolution indicators, $F_w/F$, $I_{\rm IRAC}$ and $T_c$ and their
dependence on $\Sigma$. However, we did not find any significant
correlations of these properties with $\Sigma$.

In Fig.~\ref{fig:Surfacedensity}d we also show the predictions of
eq.~(\ref{eq:LoverM}) for clumps with $M=10^3 \:M_\odot$, $T =
10,15,20~\rm K$ forming stars at fixed $\epsilon_{\rm ff}=0.002, 0.02,
0.2$.  Models with $T \sim 10-15$~K appear to define the lower
boundary of the populated region of the observed $L/M$ versus $\Sigma$
parameter space, but obtaining precise constraints on $\epsilon_{\rm
  ff}$ is difficult because of the sensitivity of $L/M$ to the adopted
temperature.  The models with high values of $\epsilon_{\rm ff}=0.2$,
even with $T =10$~K appear to exceed the observed $L/M$ of a
significant number of the clumps, thus we tentatively conclude that
$\epsilon_{\rm ff}<0.2$.  This analysis will be improved once FIR data
become available allowing individual clump temperatures to be
accurately measured from their spectral energy distributions. The
implications of the detailed distribution of $L/M$ of the clump
population and its implication for star cluster formation theories
will be examined in a future paper.



\subsection{Dependence of $L$ and $L/M$ with Velocity Dispersion and Virial Parameter}


\begin{figure}
\includegraphics[width=16cm]{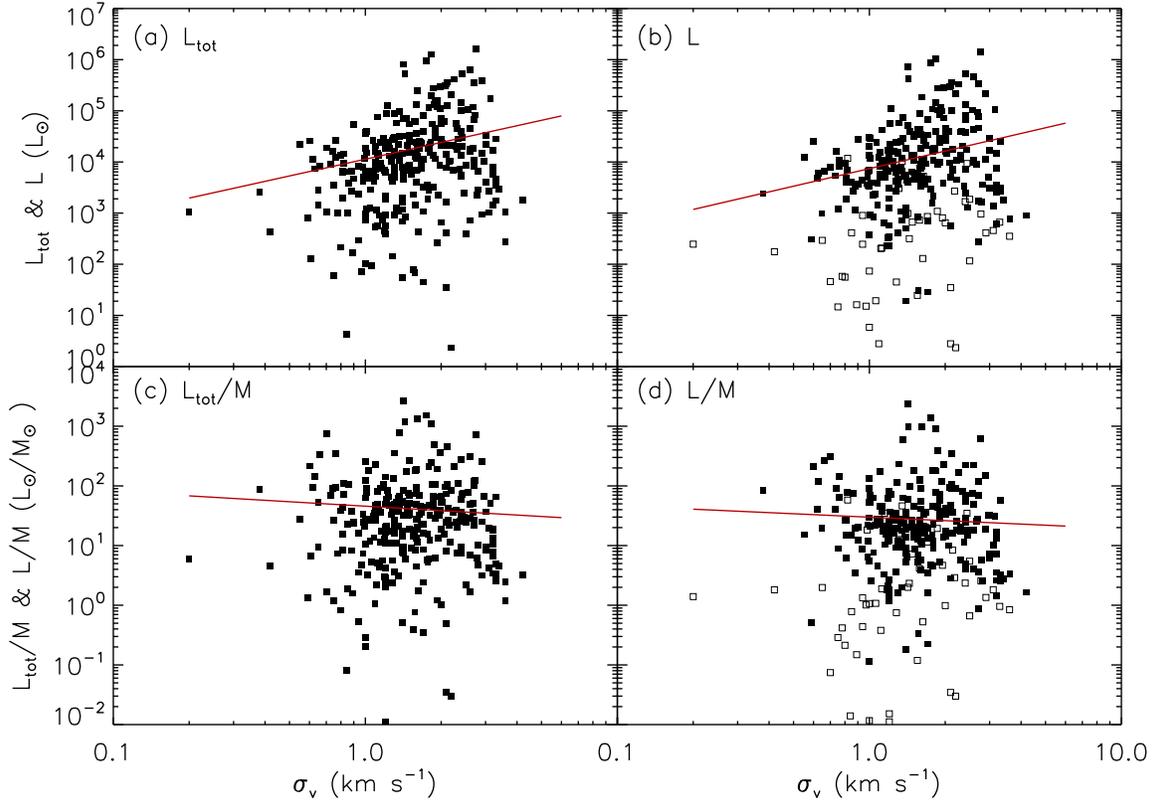}
\caption{\footnotesize 
{\it (a) Top left:} Correlation of $L_{\rm tot}$ with $\sigma_v$.
Solid line shows the best-fit relation (see text).
{\it (b) Top right:} Correlation of $L$ with $\sigma_v$.
Solid line shows the best-fit relation (see text).
{\it (c) Bottom left:} Correlation of $L_{\rm tot}/M$ with $\sigma_v$.
Solid line shows the best-fit relation (see text).
{\it (d) Bottom right:} Correlation of $L/M$ with $\sigma_v$.
Solid line shows the best-fit relation (see text).
Open squares show clumps with uncertain measurements 
 of $F$ due to IRAS 100\micron\ background subtraction (see Fig.~\ref{Ldist}b).}
\label{fig:Sigmav}
\end{figure}

\begin{figure}
\includegraphics[width=16cm]{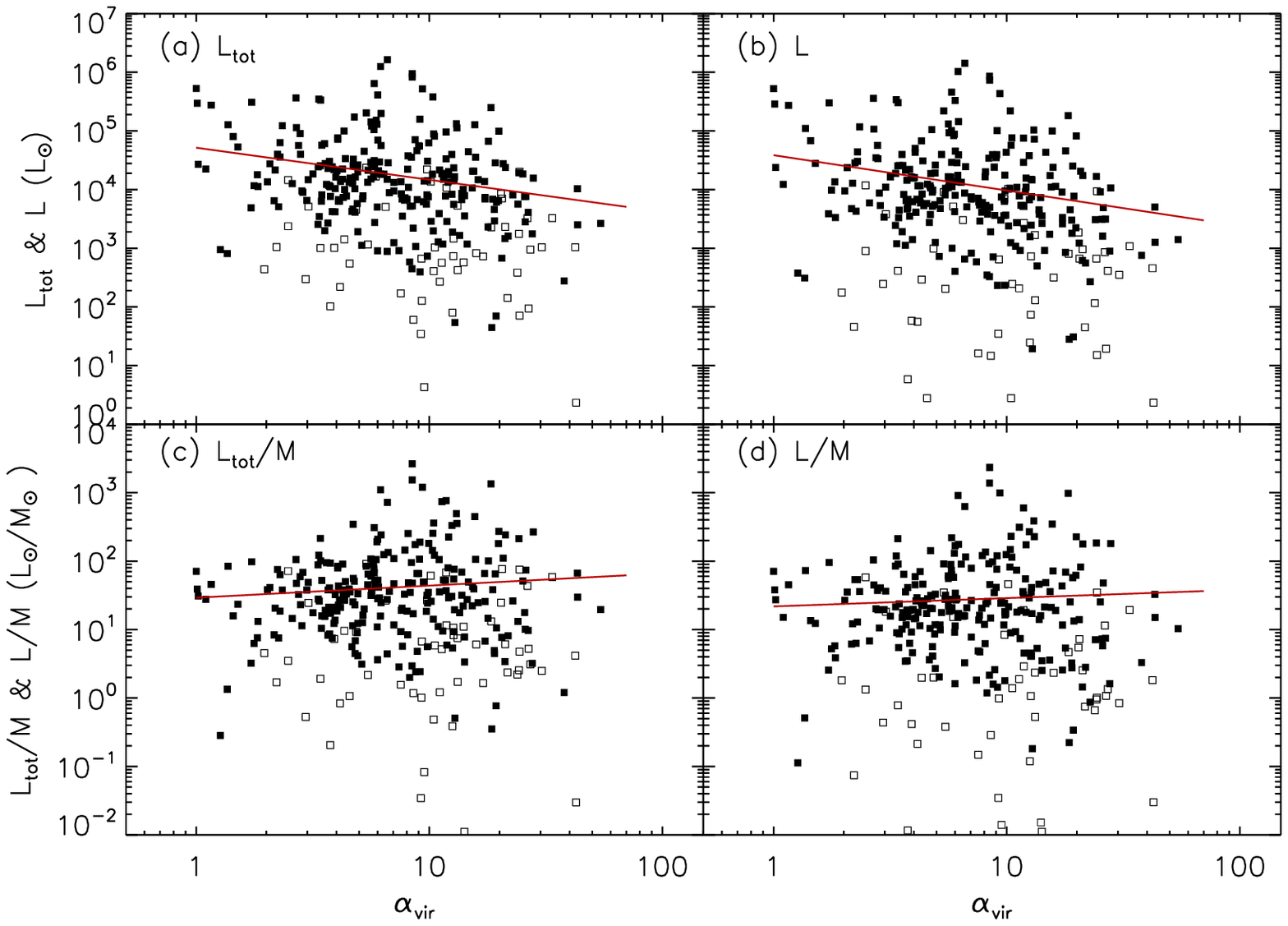}
\caption{\footnotesize 
{\it (a) Top left:} Correlation of virial parameter $\alpha$ with $L_{\rm tot}$. 
Solid line shows the best-fit relation (see text).
{\it (b) Top right:} Correlation of virial parameter $\alpha$ with $L$. 
Solid line shows the best-fit relation (see text).
{\it (c) Bottom left:} Correlation of virial parameter $\alpha$ with $L_{\rm tot}/M$. 
Solid line shows the best-fit relation (see text).
{\it (d) Bottom right:} Correlation of virial parameter $\alpha$ with $L/M$. 
Solid line shows the best-fit relation (see text). 
Open squares show clumps with uncertain measurements 
 of $F$ due to IRAS 100\micron\ background subtraction (see Fig.~\ref{Ldist}b).}
\label{fig:virial}
\end{figure}

In Fig.~\ref{fig:Sigmav}a and b, we explore the dependence of $L_{\rm
  tot}$ and $L$ on the 1D velocity dispersion, $\sigma$, (as measured from
HCO$^{+}$(1-0) in Paper I).
We find best-fit relations:
\begin{equation}
L_{\rm tot} = 11300(\pm1900)\times (\sigma/{\rm km\:s^{-1}})^{1.09\pm0.26} \:L_\odot
\end{equation}
with $r_s=0.28$ and $p_s\ll 10^{-4}$ (formally
$p_s=6.8\times10^{-6}$), and
\begin{equation}
L = 7400(\pm1200) \times (\sigma/{\rm km\:s^{-1}})^{1.14\pm0.28} \:L_\odot
\end{equation}
with $r_s=0.26$ and $p_s\lesssim 10^{-4}$ (formally
$p_s=3.2\times10^{-5}$).  We expect that $\sigma$ correlates with $M$
for clumps that are self-gravitating. Since $L$ correlates with $M$,
this can explain the observed, weaker correlation of $L$ with
$\sigma$.


Similarly, in Fig.~\ref{fig:Sigmav}c and d we show the dependence of
$L_{\rm tot}/M$ and $L/M$ with $\sigma$. We do not find significant
correlations, since the best-fit relations are:
\begin{equation}
L_{\rm tot}/M = 46(\pm6) \times (\sigma/{\rm km\:s^{-1}})^{-0.25\pm0.23}\:L_\odot/M_\odot
\end{equation}
with $r_s=-0.05$ and $p_s=0.41$, and
\begin{equation}
L/M = 30(\pm4) \times (\sigma/{\rm km\:s^{-1}})^{-0.19\pm0.24}\:L_\odot/M_\odot
\end{equation}
with $r_s=-0.04$ and $p_s=0.46$. Thus there is no apparent correlation of
these variables. If star clusters were built-up hierarchically from a
merger of smaller clumps, one might have expected to see increasing
$L/M$ with $\sigma$.

The virial parameter, $\alpha_{\rm vir}\equiv 5\sigma^2R/(GM)$
(Bertoldi \& McKee 1992), is proportional to the ratio of a clump's kinetic and
gravitational energies. In Fig.~\ref{fig:virial}a and b we show the
dependence of $L_{\rm tot}$ and $L$ with $\alpha_{\rm vir}$. 
We find best-fit relations:
\begin{equation}
L_{\rm tot} = 52000(\pm15000)\times (\alpha_{\rm vir})^{-0.55\pm0.14} \:L_\odot
\end{equation}
with $r_s=-0.24$ and $p_s=1\times10^{-4}$, and
\begin{equation}
L = 39000(\pm12000) \times (\alpha_{\rm vir})^{-0.60\pm0.15} \:L_\odot
\end{equation}
with $r_s=-0.27$ and $p_s\lesssim 10^{-4}$ (formally
$p_s=2\times10^{-5}$). These (only moderately) significant
correlations may be explained by the fact that smaller virial
parameters indicate more gravitationally bound systems, which should
be more prone to star formation. However, these relations may
alternatively be driven by the fact that more massive clumps tend to
have smaller virial parameters (Bertoldi \& McKee 1992; Paper I) and
that luminosity correlates with mass (\S\ref{S:LoverM}).

This second explanation appears to be supported by the following
results.  In Fig.~\ref{fig:virial}c and d we show the dependence of
$L_{\rm tot}/M$ and $L/M$ with $\alpha_{\rm vir}$ (note, these are
equivalent of correlating $L_{\rm tot}$ and $L$ with $\sigma^2 R)$.
We do not find significant correlations since the best-fit relations
are:
\begin{equation}
L_{\rm tot}/M = 29(\pm8) \times (\alpha_{\rm vir})^{0.18\pm0.12}\:L_\odot/M_\odot
\end{equation}
with $r_s=0.06$ and $p_s=0.36$, and
\begin{equation}
L/M = 22(\pm6) \times (\alpha_{\rm vir})^{0.12\pm0.13}\:L_\odot/M_\odot
\end{equation}
with $r_s=0.009$ and $p_s=0.89$. So these data do not reveal any correlation
of cluster evolutionary stage (as measured by $L/M$) with degree of
gravitational boundedness.

Note that the absolute values of $\alpha_{\rm vir}$ appear relatively
high, e.g. compared to the somewhat larger $\rm ^{13}CO$ clouds and
clumps analyzed by Roman-Duval et al. (2010), which have
$\bar{\alpha}_{\rm vir}\sim 1$ (see also Tan et al. 2013). As
discussed above (\S3.1), potential systematic uncertainties, especially in the
measurement of mass via an assumed HCO$^+$ abundance, may be causing
an overestimation of $\alpha_{\rm vir}$, but these uncertainties are
not expected to lead to a median value of the HCO$^+$ clump sample
that is close to unity. Thus the dynamics of the HCO$^+$ clumps may be
dominated by surface pressure, rather than by their self-gravity (see
also Paper I). This is consistent with the fact that most of the
HCO$^+$ clumps have relatively low $L/M$ and low star formation
activity, so we may expect them to have values of $\alpha_{\rm vir}$
in the range $\sim 1$ -- 30, similar to results found by Bertoldi \&
McKee (1992).

However, it is interesting that we do not see a trend of decreasing
$\alpha_{\rm vir}$ with increasing $L/M$. Possible explanations are:
(1) the uncertainties in $\alpha_{\rm vir}$ (which depends on $M$, $R$
and $\sigma^2$) and $L/M$ are large enough to wash-out any correlation
that is present; (2) the importance of self-gravity, as measured at
the HCO$^+$ clump scale, does not grow during star cluster formation.
Improved mass, luminosity and velocity dispersion measurements are
needed to investigate this issue further.

\subsection{Dependence of $L$ with HCO$^+$(1-0) line luminosity}



\citet{gao04} found a tight linear correlation between the infrared
luminosity (hereafter we refer to this as the bolometric luminosity,
$L$) and the amount of dense gas as traced by the luminosity of HCN in
both normal galaxies and starburst galaxies. This may suggest that the
star formation rate (thought to be proportional to $L$, at least in
starbursts) simply scales with the mass of dense gas. Similarly,
\citet{juneau09} found an index of $0.99\pm0.26$ in their study of the
relation between the bolometric luminosity and HCO$^+$ line luminosity
in a sample of 34 nearby galaxies.

\begin{figure}
\epsscale{1.00}
 \includegraphics[width=15cm, height=15cm]{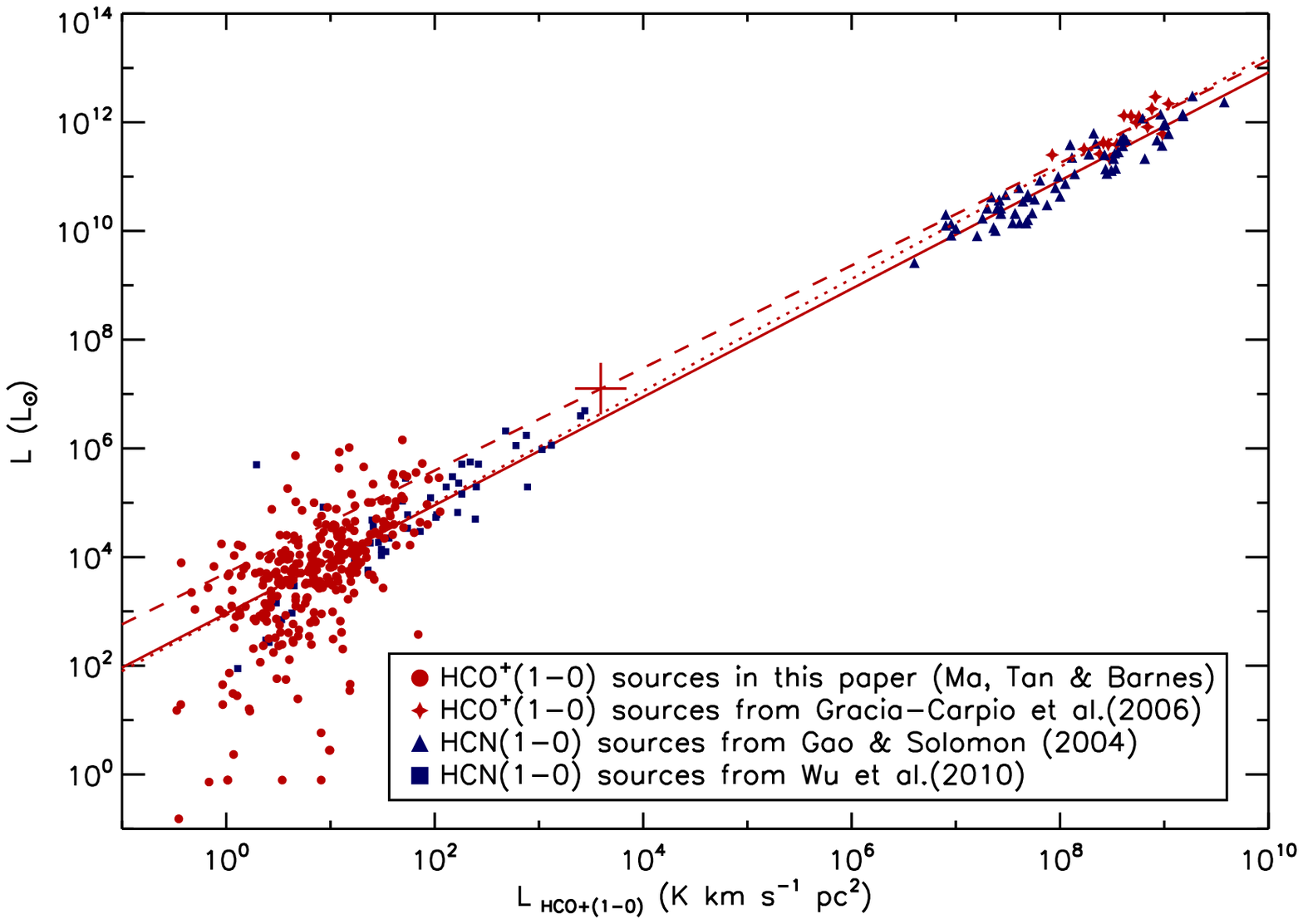}
\caption{\footnotesize 
Bolometric luminosity, $L$, versus dense gas line luminosity, $L_{\rm
  HCO^+(1-0)}$. The CHaMP clumps are shown by filled red circles. The single
large red cross shows the total luminosity and line luminosity of the
whole CHaMP sample. Other HCO$^+$(1-0) data for entire galaxies from
\citet{gracia06} are shown by red stars. We also show HCN(1-0) data of
galactic clumps (\citet{wu10} - blue squares) and entire galaxies
(\citet{gao04} - blue triangles). The best fit relation to only the CHaMP
HCO$^+$(1-0) data (filled red circles) is shown by a solid red line. The best fit relation 
to both the CHaMP (filled red circles) and extragalactic HCO$^+$(1-0) data (red stars) 
from \citet{gracia06} is shown by a dotted red line. And the best fit relation to 
the total CHaMP data point (red cross) and the extragalactic sample (red stars) 
is shown by a dashed red line.
\label{lumi-linelumi_back}}
\end{figure}

On the much smaller scales of clumps, the luminosity should not be
such a good measure of SFR (Krumholz \& Tan 2007), rather tracing
embedded stellar content. Still, by surveying a sample of massive
dense star formation clumps in CS(7-6), CS(2-1), HCN(1-0) and
HCN(3-2), Wu et al. (2005, 2010) have extended the relation of $L -
L_{\rm HCN(1-0)}$ proposed by Gao \& Solomon (2004) down to $L \sim
10^{4.5} L_{\odot}$ (see Fig.~\ref{lumi-linelumi_back}).




The CHaMP survey provides a way to connect these scales, by being a
complete census of dense gas and thus star formation activity over a
several $\rm kpc^2$ region of the Galaxy. The CHaMP clumps span the
full range of evolution of these sources that will be averaged over in
extragalactic observations. In addition, by its improved sensitivity,
the CHaMP survey allows us to extend the bolometric luminosity versus
dense gas line luminosity relation down to much smaller values of
source bolometric luminosity.

In Fig.~\ref{lumi-linelumi_back} we also plot the
CHaMP sources.  We fit a power-law relation between $L$ and $L_{\rm
  HCO^+(1-0)}$ (because of the uncertainties in background subtracted
luminosities, we only fit to those sources with
$L>10^{1.5}L_\sun$). Only fitting to the CHaMP clumps (via a least-squares
fit in log~$L$) yields:
\begin{eqnarray}
\frac{L}{L_\odot} = 917(^{+208}_{-170})  \left(\frac{L_{\rm HCO^+(1-0)}}{\rm K \; km \; s^{-1} pc^2}\right)^{1.00\pm0.09}
\label{eq:L1}
\end{eqnarray}
Similarly, a fit to both the CHaMP sample and the extragalactic
HCO$^+$(1-0) of \citet{gracia06} yields:
\begin{eqnarray}
\frac{L}{L_\odot} = 857(^{+105}_{-93} )\left(\frac{L_{\rm HCO^+(1-0)}}{\rm K \; km \; s^{-1} pc^2}\right)^{1.03\pm0.02}
\label{eq:L2}
\end{eqnarray}
Finally a fit to the total CHaMP data point and the extragalactic sample yields:
\begin{eqnarray}
\frac{L}{L_\odot} = 5100(^{+6900}_{-2900} )  \left(\frac{L_{\rm HCO^+(1-0)}}{\rm K \; km \; s^{-1} pc^2}\right)^{0.94\pm0.04}
\label{eq:L3}
\end{eqnarray}
This last fit is expected to be the most accurate for extending current extragalactic results down to lower luminosities.
Our results suggest that the $L - L_{\rm HCO^+(1-0)}$ relation in
clumps (when averaged over a complete sample) is almost the same as
that found when averaging over whole galaxies.

\section{Summary}

A total of 303 dense gas clumps have been detected using the
HCO$^+$(1$-$0) line in the CHaMP survey (Paper I). In this paper we
have derived the SED for these clumps using Spitzer, MSX and IRAS
data. By fitting a two-temperature grey-body model to the SED, we have
derived the colder component temperature, colder component flux,
warmer component temperature, warmer component flux, bolometric
temperature and bolometric flux of these dense clumps. Adopting clump
distances and $\rm HCO^+$-derived masses from Paper I, we have
calculated the bolometric luminosities and luminosity-to-mass ratios.
These dense clumps typically have 
masses $\sim 700\; M_\sun$, luminosities $\sim 5\times10^4 \; L_\sun$
and luminosity-to-mass ratios $\sim 70 \; L_\sun / M_\sun$.

During the evolution of star-forming clumps, i.e. the formation of
star clusters, the luminosity will increase and the gas mass will
decrease due to incorporation into stars and dispersal by feedback,
causing the luminosity-to-mass ratio to increase. So $L/M$ should be a
good evolutionary indicator of the star cluster formation process. The
observed range of $L/M$ from $\sim 0.1\:L_\sun / M_\sun$ to $\sim
1000\:L_\sun / M_\sun$ corresponds to that expected for evolution from
starless clumps to those with near equal mass of stars and gas.

The fraction of the warmer component flux in the bolometric flux,
$F_w/F$ has a positive correlation with the luminosity-to-mass ratio,
supporting the idea that as stars form in molecular clumps
and $L/M$ increases, a larger fraction of the bolometric flux will
come out at shorter wavelengths. We also find that the colder
component dust temperature, $T_c$, has a positive correlation with
$L/M$: the bulk of the clump material appears to be getting warmer as
star cluster formation proceeds. However, we caution that our
measurements of $T_c$ are relatively poor (they will be improved with
the acquisition of {\it Herschel} observations in this region of the
Galaxy).
We also find a highly significant correlation of specific intensity in
the {\it Spitzer}-IRAC bands (3-8~$\rm \mu m$), $I_{\rm IRAC}$ with
$L/M$. This has the potential to be a useful evolutionary indicator
for the star cluster formation process.

We investigated the dependence of $L/M$ with mass surface density,
$\Sigma$, velocity dispersion, $\sigma$ and virial parameter,
$\alpha_{\rm vir}$. The lower limit of the distribution of $L/M$ with
$\Sigma$ is consistent with a model for accretion luminosity powered
by accretion rates that are a few percent of the global clump
free-fall collapse rate. We do not see strong trends of $L/M$ with
$\Sigma$ and, if present, real effects may be masked by the intrinsic
correlation of these variables via $M$. Similarly, we do not find
strong correlations between $L/M$ and $\sigma$ or $\alpha_{\rm vir}$.


The bolometric luminosity has a nearly linear correlation with the
dense gas mass as traced by HCO$^+$(1$-$0) line luminosity, and this
relation holds for over 10 orders of magnitude from molecular clumps
in the Milky Way to infrared ultraluminous infrared galaxies. Our
results have extended the previously observed relation of \citet{wu10}
(via HCN(1$-$0) line observation) down to much lower luminosity
clumps. The complete nature of our sample also gives a measurement at
intermediate scales ($\sim$ several $\rm kpc^2$) that connects the
individual clump results with the extragalactic results, which are
averages over clump populations.






\acknowledgments JCT acknowledges support from NSF CAREER grant
AST-0645412; NASA Astrophysics Theory and Fundamental Physics grant
ATP09-0094; NASA Astrophysics Data Analysis Program ADAP10-0110.  PJB
thanks Lisa Torlina and George Papadopoulos at the University of
Sydney for their work on an earlier version of this project.

%


{\it Facilities:} \facility{IRAS}, \facility{MSX}, \facility{Spitzer (IRAC), \facility{Mopra (MOPS)}}



\appendix




\clearpage

\begin{deluxetable}{cccccccccc}
\tabletypesize{\footnotesize}
\tablecolumns{11}
\tablewidth{0pc}
\tablecaption{Primary physical properties of the HCO$^+$(1-0) clumps.$^a$ \label{tab:clumps}}
\tablehead{
\colhead{BYF} & \colhead{$\log(M)^b$} & \colhead{ $T_{\rm c, tot}$  }  &  
\colhead{$\log(F_{\rm tot})$ } & \colhead{$\log(L_{\rm tot})$ }   & 
\colhead{$\log(L_{\rm tot}/M)$} & 
\colhead{ $T_{\rm c}$  }   &\colhead{$\log(F)$}    & \colhead{$\log(L)$ }   
& \colhead{$\log(L/M)$} 
\\ 
\colhead{No.} & \colhead{$(M_\sun)$} & \colhead{ $(\rm K)$  }   & 
\colhead{$\rm erg \; s^{-1}\; cm^{-2}$ }  &  \colhead{$(L_\sun)$ }   & 
\colhead{$(L_\sun/M_\sun)$} & 
\colhead{ $(\rm K)$  }   &\colhead{$\rm erg \; s^{-1}\; cm^{-2}$ }   & \colhead{$(L_\sun)$ }   &
 \colhead{$(L_\sun/M_\sun)$}
 }  

\startdata
    5a &  3.01 &  33.4 &  -7.06 &  4.47 &  1.46 &  34.5 &  -7.07 &  4.46 &  1.45   \\
    5b &  2.13 &  30.5 &  -8.10 &  3.43 &  1.29 &  29.5 &  -8.38 &  3.15 &  1.01   \\
    5c &  2.62 &  27.8 &  -7.92 &  3.61 &  0.99 &  30.5 &  -8.03 &  3.50 &  0.87   \\
    5d &  2.95 &  34.9 &  -6.88 &  4.65 &  1.70 &  36.3 &  -6.89 &  4.64 &  1.69   \\
      7a &  3.12 &  34.8 &  -6.62 &  4.90 &  1.78 &  35.8 &  -6.64 &  4.91 &  1.79   \\
      7b &  2.71 &  34.7 &  -6.93 &  4.60 &  1.89 &  35.6 &  -6.95 &  4.58 &  1.87   \\
     9 &  2.65 &  21.5 &  -8.50 &  3.03 &  0.38 &  23.0 &  -8.68 &  2.85 &  0.20   \\
   11a &  2.90 &  37.8 &  -7.11 &  4.42 &  1.52 &  34.6 &  -7.11 &  4.42 &  1.52   \\
   11b &  2.68 &  32.4 &  -8.07 &  3.46 &  0.77 &  31.0 &  -8.19 &  3.34 &  0.66   \\
   11c &  2.10 &  30.8 &  -9.88 &  1.65 & -0.45 &  24.9 & -10.08 &  1.45 & -0.65   \\
    14$^{*}$ &  1.99 &  24.2 &  -8.89 &  2.64 &  0.65 &  26.5 &  -9.28 &  2.24 &  0.26   \\
    15$^{*}$ &  1.78 &  25.0 &  -9.38 &  2.15 &  0.37 &  30.5 &  -9.88 &  1.65 & -0.13   \\
   16a &  1.46 &  41.8 &  -8.12 &  3.41 &  1.95 &  41.1 &  -8.14 &  3.39 &  1.93   \\
   16b &  2.30 &  25.8 &  -8.58 &  2.95 &  0.64 &  22.2 &  -8.71 &  2.82 &  0.51   \\
   16c$^{*}$ &  2.04 &  25.2 &  -8.78 &  2.75 &  0.71 &  29.2 &  -9.21 &  2.32 &  0.28   \\
   16d &  2.06 &  32.3 &  -8.42 &  3.11 &  1.04 &  34.2 &  -8.50 &  3.03 &  0.97   \\
   17a &  2.11 &  31.0 &  -8.19 &  3.34 &  1.23 &  36.0 &  -8.25 &  3.28 &  1.17   \\
   17b &  2.09 &  26.7 &  -8.46 &  3.07 &  0.98 &  29.0 &  -8.57 &  2.96 &  0.87   \\
   17c &  2.55 &  25.7 &  -8.48 &  3.05 &  0.50 &  27.2 &  -8.67 &  2.86 &  0.31   \\
   19a$^{*}$ &  2.13 &  27.4 &  -8.61 &  2.92 &  0.78 &  36.7 &  -9.03 &  2.50 &  0.37   \\
   19b &  1.94 &  21.5 &  -8.70 &  2.83 &  0.89 &  12.8 &  -8.44 &  3.09 &  1.14   \\
    20$^{*}$ &  2.39 &  24.5 &  -8.90 &  2.63 &  0.24 &  29.2 &  -9.42 &  2.11 & -0.28   \\
    22 &  2.66 &  27.9 &  -7.83 &  3.70 &  1.04 &  32.2 &  -7.91 &  3.62 &  0.96   \\
    23a &  2.65 &  30.4 &  -7.65 &  3.88 &  1.24 &  29.0 &  -7.70 &  3.83 &  1.19   \\
  23b &  2.52 & 30.9 &  -7.39 & 4.14 &  1.62 & 30.5 & -7.41 & 4.12 &  1.60 \\
    24 &  2.57 &  29.3 &  -7.53 &  4.00 &  1.43 &  30.8 &  -7.60 &  3.92 &  1.35   \\
    25a &  3.06 &  29.7 &  -7.14 &  4.39 &  1.33 &  34.0 &  -7.19 &  4.34 &  1.28   \\
  25b &  1.95 &  30.2 &  -8.24 & 3.29 & 1.34 &  34.0 &  -8.38 & 3.15 & 1.20   \\
    26 &  2.49 &  28.0 &  -7.77 &  3.75 &  1.27 &  32.5 &  -7.97 &  3.56 &  1.07   \\
    27$^{*}$ &  2.17 &  26.7 &  -8.38 &  3.15 &  0.98 &  30.7 &  -9.06 &  2.47 &  0.30   \\
   32a &  2.40 &  28.9 &  -8.57 &  2.96 &  0.56 &  29.0 &  -8.92 &  2.61 &  0.21   \\
   32b &  2.12 &  26.9 &  -8.66 &  2.87 &  0.75 &  31.9 &  -9.16 &  2.37 &  0.25   \\
   36a &  2.68 &  30.0 &  -7.64 &  3.89 &  1.21 &  31.0 &  -7.85 &  3.68 &  1.00   \\
   36b &  2.81 &  33.6 &  -7.39 &  4.14 &  1.33 &  31.4 &  -7.46 &  4.07 &  1.25   \\
   36c &  2.66 &  35.3 &  -7.03 &  4.50 &  1.83 &  36.6 &  -7.07 &  4.46 &  1.80   \\
   36d &  2.65 &  29.9 &  -7.60 &  3.93 &  1.28 &  30.2 &  -7.74 &  3.79 &  1.14   \\
   36e &  2.20 &  29.5 &  -8.17 &  3.36 &  1.17 &  24.1 &  -8.60 &  2.93 &  0.73   \\
   37a$^{*}$ &  2.14 &  28.3 &  -8.52 &  3.01 &  0.86 &  50.0 &  -9.77 &  1.76 & -0.38   \\
   37b$^{*}$ &  1.84 &  29.3 &  -8.66 &  2.87 &  1.03 &  41.4 &  -9.66 &  1.87 &  0.03   \\
    38 &  2.10 &  35.0 &  -7.82 &  3.30 &  1.20 &  35.3 &  -7.84 &  3.28 &  1.18   \\
   40a &  3.88 &  42.7 &  -6.69 &  5.47 &  1.59 &  41.7 &  -6.70 &  5.46 &  1.58   \\
   40b &  3.70 &  34.1 &  -7.25 &  4.90 &  1.20 &  34.3 &  -7.32 &  4.83 &  1.13   \\
   40c &  3.02 &  32.9 &  -7.82 &  4.33 &  1.31 &  34.9 &  -8.05 &  4.11 &  1.08   \\
   40d &  3.38 &  34.5 &  -7.10 &  5.06 &  1.68 &  33.9 &  -7.13 &  5.03 &  1.65   \\
   40e &  3.53 &  33.2 &  -7.36 &  4.80 &  1.27 &  30.7 &  -7.52 &  4.64 &  1.11   \\
   40f &  3.59 &  31.9 &  -7.37 &  4.78 &  1.19 &  30.4 &  -7.56 &  4.60 &  1.01   \\
   40g &  3.36 &  32.6 &  -7.41 &  4.75 &  1.39 &  30.3 &  -7.57 &  4.59 &  1.23   \\
    41 &  3.36 &  28.0 &  -7.43 &  4.73 &  1.37 &  30.7 &  -7.71 &  4.45 &  1.09   \\
   42a &  2.74 &  28.7 &  -8.09 &  4.07 &  1.33 &  29.3 &  -8.35 &  3.81 &  1.07   \\
   42b &  2.78 &  30.0 &  -8.35 &  3.80 &  1.02 &  30.8 &  -8.64 &  3.52 &  0.74   \\
    47$^{*}$ &  2.83 &  26.6 &  -8.59 &  3.37 &  0.54 &  28.8 &  -9.01 &  2.95 &  0.12   \\
   54a &  3.50 &  41.2 &  -6.48 &  5.49 &  1.99 &  47.0 &  -6.49 &  5.48 &  1.98   \\
   54b &  3.78 &  39.5 &  -6.53 &  5.44 &  1.66 &  40.7 &  -6.53 &  5.43 &  1.65   \\
   54c &  3.51 &  32.9 &  -6.88 &  5.09 &  1.58 &  28.5 &  -6.90 &  5.07 &  1.56   \\
   54d &  3.54 &  42.6 &  -6.40 &  5.56 &  2.02 &  43.7 &  -6.41 &  5.56 &  2.02   \\
   54e &  3.37 &  32.5 &  -6.81 &  5.15 &  1.78 &  37.6 &  -6.85 &  5.12 &  1.75   \\
   54f &  3.25 &  32.7 &  -7.63 &  4.34 &  1.09 &  36.1 &  -7.75 &  4.21 &  0.97   \\
   54g &  3.15 &  25.2 &  -8.14 &  3.83 &  0.68 &  27.0 &  -8.41 &  3.56 &  0.41   \\
   54h &  3.17 &  32.1 &  -7.83 &  4.14 &  0.97 &  34.3 &  -8.11 &  3.86 &  0.69   \\
   56a &  3.35 &  35.6 &  -7.32 &  4.65 &  1.30 &  37.2 &  -7.37 &  4.60 &  1.25   \\
   56b &  3.02 &  26.3 &  -8.42 &  3.55 &  0.53 &  24.4 &  -8.54 &  3.42 &  0.40   \\
   56c$^{*}$ &  2.84 &  22.8 &  -8.72 &  3.25 &  0.40 &  21.5 &  -9.14 &  2.82 & -0.02   \\
   56d &  2.97 &  29.2 &  -8.31 &  3.66 &  0.69 &  28.7 &  -8.49 &  3.48 &  0.51   \\
   57a &  3.05 &  27.5 &  -7.86 &  4.10 &  1.06 &  31.0 &  -7.90 &  4.06 &  1.02   \\
   57b &  2.54 &  21.2 &  -8.69 &  3.27 &  0.73 &  16.9 &  -8.77 &  3.20 &  0.65   \\
   60a &  2.51 &  22.5 &  -8.54 &  3.43 &  0.92 &  24.4 &  -8.92 &  3.05 &  0.54   \\
   60b$^{*}$ &  2.25 &  21.6 &  -8.95 &  3.02 &  0.77 &  26.4 &  -9.57 &  2.39 &  0.14   \\
   61a$^{*}$ &  2.48 &  26.2 &  -8.34 &  2.98 &  0.49 &  27.3 &  -8.70 &  2.61 &  0.13   \\
   61b$^{*}$ &  2.40 &  27.8 &  -8.30 &  3.02 &  0.62 &  25.5 &  -8.65 &  2.66 &  0.26   \\
    62 &  2.36 &  14.5 &  -8.87 &  2.44 &  0.08 &  12.8 &  -8.44 &  2.88 &  0.52   \\
    63$^{*}$ &  2.24 &  26.1 &  -8.73 &  2.59 &  0.34 &  28.0 &  -9.25 &  2.07 & -0.18   \\
    66 &  2.49 &  25.4 &  -7.40 &  3.91 &  1.42 &  23.5 &  -8.88 &  2.43 & -0.06   \\
    67 &  2.63 &  24.0 &  -8.15 &  3.17 &  0.53 &  22.8 &  -8.52 &  2.79 &  0.16   \\
    68 &  3.43 &  24.4 &  -6.97 &  4.35 &  0.92 &  29.6 &  -7.10 &  4.22 &  0.79   \\
    69 &  2.55 &  30.3 &  -7.18 &  4.13 &  1.58 &  30.3 &  -7.34 &  3.98 &  1.43   \\
   70a &  2.94 &  34.7 &  -6.93 &  4.39 &  1.44 &  35.8 &  -7.04 &  4.28 &  1.33   \\
   70b &  3.05 &  35.3 &  -6.99 &  4.32 &  1.27 &  37.6 &  -7.11 &  4.21 &  1.16   \\
    71 &  2.63 &  30.5 &  -7.43 &  3.88 &  1.25 &  31.1 &  -7.82 &  3.49 &  0.86   \\
    72 &  3.38 &  27.5 &  -7.10 &  4.22 &  0.84 &  31.7 &  -7.32 &  3.99 &  0.61   \\
    73 &  3.16 &  31.3 &  -6.87 &  4.44 &  1.28 &  34.8 &  -6.90 &  4.41 &  1.25   \\
    76 &  2.76 &  24.5 &  -7.53 &  3.78 &  1.03 &  21.4 &  -7.43 &  3.89 &  1.13   \\
   77a &  2.93 &  28.1 &  -7.26 &  4.05 &  1.12 &  29.5 &  -7.79 &  3.52 &  0.59   \\
   77b &  3.27 &  33.2 &  -7.25 &  4.07 &  0.80 &  31.5 &  -7.32 &  3.99 &  0.72   \\
   77c &  3.24 &  33.1 &  -6.71 &  4.60 &  1.36 &  36.0 &  -6.79 &  4.53 &  1.29   \\
   77d &  2.68 &  34.5 &  -7.13 &  4.18 &  1.50 &  33.5 &  -7.28 &  4.03 &  1.35   \\
   78a &  2.68 &  26.7 &  -7.29 &  4.03 &  1.35 &  28.0 &  -8.08 &  3.24 &  0.56   \\
   78b &  3.01 &  26.7 &  -7.42 &  3.90 &  0.89 &  28.5 &  -7.76 &  3.55 &  0.55   \\
   78c$^{*}$ &  2.62 &  23.2 &  -8.30 &  3.02 &  0.40 &  23.8 &  -8.77 &  2.55 & -0.07   \\
   79a &  2.82 &  31.4 &  -7.85 &  3.47 &  0.65 &  29.6 &  -8.06 &  3.26 &  0.44   \\
   79b &  2.63 &  31.3 &  -7.56 &  3.75 &  1.12 &  31.3 &  -7.83 &  3.49 &  0.85   \\
   79c &  2.74 &  29.6 &  -8.07 &  3.25 &  0.51 &  29.3 &  -8.36 &  2.95 &  0.21   \\
    83 &  2.74 &  32.6 &  -7.14 &  4.18 &  1.44 &  32.0 &  -7.37 &  3.94 &  1.20   \\
   85a &  2.96 &  26.6 &  -7.70 &  3.61 &  0.65 &  19.6 &  -7.60 &  3.72 &  0.75   \\
   85b &  3.01 &  30.1 &  -7.48 &  3.84 &  0.83 &  28.5 &  -7.90 &  3.42 &  0.41   \\
   85c &  2.30 &  29.6 &  -8.11 &  3.21 &  0.91 &  25.1 &  -8.57 &  2.75 &  0.45   \\
   86a &  2.78 &  30.5 &  -7.57 &  3.75 &  0.97 &  34.1 &  -7.92 &  3.39 &  0.61   \\
   86b &  2.57 &  31.9 &  -7.48 &  3.84 &  1.27 &  33.3 &  -7.67 &  3.65 &  1.08   \\
    87 &  3.12 &  29.4 &  -7.26 &  4.06 &  0.94 &  27.2 &  -7.65 &  3.67 &  0.54   \\
    88 &  2.87 &  35.2 &  -6.91 &  4.41 &  1.54 &  37.6 &  -7.18 &  4.14 &  1.27   \\
   89a &  2.45 &  30.2 &  -7.86 &  3.45 &  1.01 &  26.0 &  -8.11 &  3.21 &  0.76   \\
   89b &  3.07 &  31.6 &  -7.05 &  4.26 &  1.20 &  25.3 &  -7.07 &  4.24 &  1.18   \\
   89c &  2.14 &  30.4 &  -8.02 &  3.30 &  1.15 &  27.0 &  -8.22 &  3.10 &  0.95   \\
   90a &  2.87 &  32.7 &  -6.83 &  4.48 &  1.61 &  32.0 &  -7.16 &  4.16 &  1.29   \\
   90b &  2.78 &  33.8 &  -6.98 &  4.33 &  1.55 &  32.9 &  -7.28 &  4.04 &  1.26   \\
   90c &  2.53 &  34.5 &  -6.76 &  4.55 &  2.03 &  34.6 &  -6.93 &  4.38 &  1.86   \\
   91a &  3.24 &  31.2 &  -6.56 &  4.75 &  1.51 &  35.0 &  -6.77 &  4.54 &  1.30   \\
   91b &  2.43 &  31.7 &  -7.38 &  3.94 &  1.50 &  29.8 &  -7.73 &  3.58 &  1.15   \\
   91c$^{*}$ &  2.58 &  28.2 &  -7.95 &  3.36 &  0.79 &  19.5 &  -8.33 &  2.98 &  0.41   \\
   91d$^{*}$ &  2.82 &  27.9 &  -8.49 &  2.82 &  0.01 &  11.9 &  -8.50 &  2.81 & -0.01   \\
   91e &  2.53 &  33.8 &  -7.19 &  4.12 &  1.60 &  29.1 &  -7.55 &  3.76 &  1.23   \\
   92a$^{*}$ &  2.85 &  28.1 &  -7.59 &  3.73 &  0.87 &  29.6 &  -8.09 &  3.22 &  0.37   \\
   92b$^{*}$ &  2.97 &  30.0 &  -7.28 &  4.03 &  1.06 &  26.5 &  -7.88 &  3.43 &  0.46   \\
   93a &  3.13 &  34.6 &  -6.47 &  4.85 &  1.72 &  37.8 &  -6.66 &  4.66 &  1.53   \\
   93b &  2.82 &  35.5 &  -6.49 &  4.82 &  2.00 &  38.5 &  -6.66 &  4.66 &  1.84   \\
   93c &  2.56 &  34.3 &  -7.05 &  4.27 &  1.71 &  31.8 &  -7.35 &  3.96 &  1.40   \\
   94a &  2.98 &  34.3 &  -6.83 &  4.48 &  1.50 &  33.2 &  -7.07 &  4.25 &  1.26   \\
   94b &  2.85 &  31.6 &  -7.02 &  4.30 &  1.45 &  29.3 &  -7.41 &  3.90 &  1.05   \\
   94c &  2.79 &  33.6 &  -7.06 &  4.25 &  1.46 &  32.9 &  -7.28 &  4.04 &  1.25   \\
   94d &  2.84 &  33.8 &  -6.65 &  4.66 &  1.82 &  34.4 &  -6.93 &  4.38 &  1.54   \\
   94e &  2.19 &  35.1 &  -7.30 &  4.02 &  1.83 &  33.5 &  -7.61 &  3.70 &  1.51   \\
   94f &  2.01 &  34.8 &  -7.43 &  3.88 &  1.87 &  33.4 &  -7.62 &  3.69 &  1.68   \\
   94g &  2.40 &  34.6 &  -6.87 &  4.44 &  2.04 &  34.9 &  -7.00 &  4.31 &  1.92   \\
   94h &  2.56 &  35.4 &  -6.32 &  5.00 &  2.44 &  41.6 &  -6.40 &  4.92 &  2.36   \\
   95a &  2.94 &  35.2 &  -7.21 &  4.10 &  1.16 &  32.7 &  -7.54 &  3.77 &  0.83   \\
   95b &  2.17 &  35.2 &  -6.74 &  4.57 &  2.40 &  35.3 &  -7.01 &  4.30 &  2.13   \\
   95c$^{*}$ &  2.28 &  34.1 &  -7.61 &  3.70 &  1.42 &  14.4 &  -7.83 &  3.48 &  1.20   \\
    96 &  2.81 &  30.0 &  -7.55 &  3.76 &  0.95 &  19.7 &  -7.57 &  3.74 &  0.94   \\
    97 &  2.77 &  33.0 &  -6.80 &  4.52 &  1.75 &  34.1 &  -7.04 &  4.27 &  1.50   \\
   98a &  2.93 &  39.6 &  -6.11 &  5.20 &  2.28 &  44.7 &  -6.31 &  5.00 &  2.08   \\
   98b$^{*}$ &  2.53 &  39.6 &  -7.39 &  3.93 &  1.39 &  46.1 &  -8.04 &  3.27 &  0.74   \\
   98c$^{*}$ &  1.89 &  45.1 & -10.95 &  0.37 & -1.53 &  10.0 & -10.95 &  0.37 & -1.52   \\
   99a &  2.83 &  35.5 &  -6.87 &  4.44 &  1.61 &  35.7 &  -6.94 &  4.37 &  1.55   \\
   99b &  2.75 &  35.5 &  -6.76 &  4.56 &  1.80 &  39.3 &  -6.83 &  4.48 &  1.73   \\
   99c &  2.33 &  41.8 &  -6.86 &  4.46 &  2.13 &  41.7 &  -6.92 &  4.39 &  2.06   \\
   99d &  1.88 &  38.2 &  -6.90 &  4.41 &  2.54 &  32.2 &  -7.29 &  4.03 &  2.15   \\
   99e &  2.41 &  38.1 &  -7.40 &  3.91 &  1.50 &  37.5 &  -7.51 &  3.80 &  1.39   \\
   99f &  1.99 &  42.9 &  -6.94 &  4.38 &  2.38 &  50.0 &  -7.07 &  4.24 &  2.25   \\
   99g &  2.46 &  46.6 &  -6.21 &  5.11 &  2.65 &  50.0 &  -6.32 &  5.00 &  2.54   \\
   99h$^{*}$ &  2.31 &  43.1 &  -7.15 &  4.16 &  1.85 &  47.4 &  -7.24 &  4.07 &  1.76   \\
   99i &  2.10 &  43.5 &  -6.33 &  4.98 &  2.88 &  45.2 &  -6.44 &  4.88 &  2.78   \\
   99j &  3.18 &  41.0 &  -6.21 &  5.11 &  1.93 &  42.9 &  -6.27 &  5.04 &  1.86   \\
   99k &  2.91 &  40.1 &  -6.02 &  5.30 &  2.38 &  47.0 &  -6.16 &  5.16 &  2.25   \\
   99l &  3.44 &  36.4 &  -6.32 &  5.00 &  1.56 &  40.2 &  -6.48 &  4.83 &  1.39   \\
   99m &  3.52 &  39.4 &  -5.79 &  5.53 &  2.00 &  50.0 &  -5.83 &  5.48 &  1.96   \\
   99n &  2.71 &  36.4 &  -6.57 &  4.74 &  2.03 &  37.5 &  -6.72 &  4.59 &  1.88   \\
   99o &  2.46 &  33.4 &  -7.17 &  4.15 &  1.69 &  32.0 &  -7.44 &  3.88 &  1.42   \\
   99p &  2.11 &  38.3 &  -7.05 &  4.26 &  2.16 &  41.9 &  -7.29 &  4.02 &  1.91   \\
   99q &  2.51 &  36.8 &  -7.30 &  4.02 &  1.51 &  38.8 &  -7.52 &  3.79 &  1.29   \\
   99r &  2.61 &  33.4 &  -6.46 &  4.85 &  2.24 &  37.7 &  -6.72 &  4.60 &  1.99   \\
  100a &  3.04 &  34.2 &  -6.56 &  4.76 &  1.72 &  34.8 &  -6.90 &  4.41 &  1.37   \\
  100b &  2.83 &  37.6 &  -6.20 &  5.11 &  2.28 &  40.8 &  -6.37 &  4.95 &  2.11   \\
  100c &  1.96 &  39.7 &  -7.03 &  4.29 &  2.33 &  37.1 &  -7.09 &  4.22 &  2.27   \\
  100d$^{*}$ &  2.53 &  34.8 &  -7.09 &  4.22 &  1.70 &  32.2 &  -7.60 &  3.71 &  1.18   \\
  100e$^{*}$ &  2.43 &  36.0 &  -7.05 &  4.27 &  1.83 &  38.7 &  -7.60 &  3.72 &  1.28   \\
  100f &  1.97 &  31.8 &  -8.09 &  3.22 &  1.25 &  28.9 &  -8.62 &  2.70 &  0.72   \\
  100g$^{*}$ &  2.25 &  32.0 &  -8.15 &  3.16 &  0.92 &  12.8 &  -8.45 &  2.86 &  0.62   \\
  101a &  2.17 &  35.6 &  -7.42 &  3.89 &  1.72 &  34.0 &  -7.59 &  3.73 &  1.55   \\
  101b &  2.32 &  36.4 &  -7.18 &  4.14 &  1.82 &  35.5 &  -7.36 &  3.96 &  1.64   \\
  102a &  2.24 &  38.7 &  -7.50 &  3.82 &  1.58 &  38.1 &  -7.79 &  3.53 &  1.29   \\
  102b &  2.43 &  38.6 &  -6.85 &  4.47 &  2.03 &  36.2 &  -7.10 &  4.22 &  1.78   \\
  102c$^{*}$ &  1.75 &  37.4 &  -7.80 &  3.52 &  1.77 &  41.5 &  -8.28 &  3.04 &  1.29   \\
  102d &  1.56 &  40.0 &  -7.50 &  3.81 &  2.26 &  41.1 &  -7.97 &  3.35 &  1.79   \\
  103a$^{*}$ &  2.41 &  38.8 &  -6.94 &  4.37 &  1.96 &  48.2 &  -7.36 &  3.96 &  1.54   \\
  103b$^{*}$ &  2.43 &  39.5 &  -7.01 &  4.30 &  1.88 &  42.5 &  -7.35 &  3.97 &  1.54   \\
  103c &  2.39 &  38.2 &  -6.36 &  4.95 &  2.56 &  40.5 &  -6.56 &  4.75 &  2.36   \\
  103d &  2.48 &  39.0 &  -7.15 &  4.17 &  1.69 &  37.7 &  -7.27 &  4.04 &  1.56   \\
  103e &  1.77 &  38.4 &  -7.12 &  4.20 &  2.43 &  35.8 &  -7.29 &  4.03 &  2.26   \\
  104a &  1.40 &  42.1 &  -7.05 &  4.27 &  2.87 &  32.8 &  -7.42 &  3.89 &  2.50   \\
  104b &  2.50 &  44.9 &  -6.26 &  5.05 &  2.55 &  47.6 &  -6.46 &  4.86 &  2.36   \\
  104c &  2.43 &  43.5 &  -6.19 &  5.12 &  2.69 &  46.7 &  -6.30 &  5.02 &  2.59   \\
  105a &  2.37 &  34.6 &  -7.51 &  3.80 &  1.43 &  21.1 &  -7.48 &  3.83 &  1.47   \\
  105b &  3.16 &  35.9 &  -6.13 &  5.18 &  2.02 &  40.3 &  -6.39 &  4.92 &  1.76   \\
  105c$^{*}$ &  2.64 &  29.2 &  -8.45 &  2.86 &  0.22 &  35.6 & -11.42 & -0.10 & -2.74   \\
  105d$^{*}$ &  2.92 &  35.2 &  -6.97 &  4.34 &  1.43 &  26.0 &  -7.48 &  3.84 &  0.92   \\
  105e &  2.27 &  35.2 &  -7.17 &  4.15 &  1.88 &  31.8 &  -7.43 &  3.88 &  1.61   \\
  106a &  2.47 &  36.0 &  -7.19 &  4.13 &  1.66 &  34.2 &  -7.40 &  3.91 &  1.44   \\
  106b &  2.22 &  38.3 &  -7.19 &  4.12 &  1.90 &  37.0 &  -7.35 &  3.96 &  1.74   \\
  106c$^{*}$ &  1.72 &  37.7 &  -8.56 &  2.76 &  1.04 &  35.6 & -11.42 & -0.10 & -1.82   \\
  107a &  2.22 &  37.4 &  -7.14 &  4.18 &  1.96 &  36.2 &  -7.25 &  4.07 &  1.85   \\
  107b &  2.72 &  36.2 &  -6.71 &  4.60 &  1.88 &  36.9 &  -6.86 &  4.45 &  1.73   \\
  107c &  2.54 &  34.8 &  -6.52 &  4.80 &  2.26 &  36.5 &  -6.68 &  4.64 &  2.10   \\
  107d &  2.17 &  34.7 &  -7.05 &  4.26 &  2.10 &  32.5 &  -7.19 &  4.12 &  1.96   \\
  107e &  2.73 &  36.9 &  -6.93 &  4.39 &  1.66 &  36.3 &  -7.26 &  4.05 &  1.32   \\
  107f &  2.63 &  35.8 &  -6.95 &  4.37 &  1.74 &  32.7 &  -7.14 &  4.18 &  1.55   \\
  107g &  2.52 &  36.4 &  -6.98 &  4.33 &  1.81 &  31.7 &  -7.29 &  4.02 &  1.50   \\
  107h$^{*}$ &  2.35 &  33.3 &  -7.18 &  4.14 &  1.79 &  23.2 &  -7.30 &  4.01 &  1.66   \\
  107i$^{*}$ &  1.88 &  36.3 &  -7.80 &  3.51 &  1.64 &  25.8 &  -8.38 &  2.93 &  1.06   \\
  108a &  2.49 &  34.7 &  -7.69 &  3.62 &  1.14 &  29.7 &  -8.17 &  3.15 &  0.66   \\
  108b &  2.93 &  34.0 &  -6.57 &  4.75 &  1.82 &  36.3 &  -6.77 &  4.54 &  1.62   \\
  108c$^{*}$ &  2.24 &  35.1 &  -7.96 &  3.36 &  1.12 &  35.3 &  -8.41 &  2.91 &  0.67   \\
  108d$^{*}$ &  1.96 &  33.8 &  -7.47 &  3.85 &  1.89 &  26.5 &  -8.49 &  2.82 &  0.86   \\
  109a &  2.94 &  42.0 &  -6.88 &  4.43 &  1.49 &  39.4 &  -6.94 &  4.38 &  1.44   \\
  109b &  2.35 &  34.4 &  -7.45 &  3.87 &  1.52 &  35.6 &  -7.54 &  3.77 &  1.42   \\
  109c &  2.33 &  34.7 &  -7.44 &  3.88 &  1.54 &  30.5 &  -7.75 &  3.56 &  1.23   \\
  109d &  1.99 &  33.4 &  -7.48 &  3.83 &  1.84 &  27.3 &  -7.66 &  3.65 &  1.66   \\
  109e &  1.40 &  38.2 &  -7.40 &  3.92 &  2.52 &  38.9 &  -7.49 &  3.83 &  2.43   \\
  109f$^{*}$ &  2.32 &  38.1 &  -7.60 &  3.71 &  1.39 &  18.1 &  -7.73 &  3.58 &  1.26   \\
  110a &  2.93 &  35.1 &  -6.46 &  4.86 &  1.93 &  39.4 &  -6.61 &  4.70 &  1.77   \\
  110b &  2.12 &  36.9 &  -7.36 &  3.95 &  1.84 &  36.4 &  -7.59 &  3.73 &  1.61   \\
  111a &  3.53 &  33.1 &  -6.23 &  5.08 &  1.56 &  39.0 &  -6.35 &  4.97 &  1.44   \\
  111b &  2.25 &  36.3 &  -7.48 &  3.84 &  1.58 &  35.9 &  -7.66 &  3.66 &  1.40   \\
  111c &  1.92 &  33.8 &  -7.92 &  3.40 &  1.47 &  31.9 &  -8.21 &  3.10 &  1.18   \\
  111d &  2.72 &  29.2 &  -7.00 &  4.31 &  1.59 &  30.5 &  -7.27 &  4.04 &  1.32   \\
   112$^{*}$ &  2.69 &  31.0 &  -7.66 &  3.66 &  0.97 &  26.9 &  -8.32 &  2.99 &  0.30   \\
  113a &  2.11 &  38.4 &  -7.23 &  4.09 &  1.98 &  38.3 &  -7.40 &  3.92 &  1.81   \\
  113b &  1.96 &  39.5 &  -7.01 &  4.31 &  2.35 &  38.7 &  -7.12 &  4.20 &  2.24   \\
  114a &  2.35 &  32.9 &  -7.28 &  4.04 &  1.69 &  31.0 &  -7.58 &  3.73 &  1.38   \\
  114b &  2.30 &  37.5 &  -7.50 &  3.82 &  1.52 &  31.7 &  -7.70 &  3.61 &  1.31   \\
  114c &  2.08 &  34.2 &  -7.27 &  4.05 &  1.97 &  35.0 &  -7.62 &  3.69 &  1.61   \\
  115a$^{*}$ &  2.34 &  32.7 &  -8.89 &  2.43 &  0.08 &  35.6 & -11.42 & -0.10 & -2.45   \\
  115b &  1.94 &  35.7 &  -7.34 &  3.98 &  2.04 &  35.7 &  -7.48 &  3.84 &  1.90   \\
  115c &  1.57 &  33.8 &  -7.72 &  3.59 &  2.02 &  34.0 &  -7.88 &  3.44 &  1.87   \\
  116a &  2.12 &  33.6 &  -7.77 &  3.54 &  1.42 &  32.7 &  -8.24 &  3.08 &  0.95   \\
  116b &  1.72 &  33.7 &  -7.87 &  3.44 &  1.73 &  32.6 &  -8.31 &  3.00 &  1.29   \\
  116c &  2.10 &  34.3 &  -7.52 &  3.79 &  1.70 &  34.7 &  -7.83 &  3.48 &  1.38   \\
  117a &  2.77 &  33.0 &  -6.96 &  4.35 &  1.58 &  30.7 &  -7.49 &  3.83 &  1.05   \\
  117b &  2.46 &  34.8 &  -7.45 &  3.87 &  1.41 &  32.7 &  -7.79 &  3.52 &  1.07   \\
  117c$^{*}$ &  1.85 &  30.6 & -11.42 & -0.10 & -1.95 &  35.6 & -11.42 & -0.10 & -1.95   \\
  117d &  2.19 &  33.8 &  -7.48 &  3.83 &  1.64 &  32.4 &  -7.70 &  3.61 &  1.42   \\
  117e &  2.45 &  34.3 &  -7.19 &  4.12 &  1.67 &  32.7 &  -7.44 &  3.88 &  1.42   \\
  118a &  2.83 &  32.3 &  -7.61 &  3.71 &  0.88 &  35.0 &  -7.68 &  3.63 &  0.80   \\
  118b &  2.39 &  33.4 &  -7.37 &  3.95 &  1.56 &  31.2 &  -7.61 &  3.71 &  1.32   \\
  118c &  2.70 &  34.5 &  -6.99 &  4.32 &  1.62 &  33.5 &  -7.18 &  4.13 &  1.43   \\
  123a &  2.47 &  25.2 &  -9.42 &  2.77 &  0.30 &  23.2 &  -9.64 &  2.54 &  0.08   \\
  123b$^{*}$ &  2.75 &  26.1 &  -9.71 &  2.47 & -0.28 &  13.1 &  -9.79 &  2.39 & -0.36   \\
  123c$^{*}$ &  2.70 &  23.8 & -10.17 &  2.01 & -0.69 &  35.6 & -11.42 &  0.77 & -1.93   \\
  123d$^{*}$ &  2.42 &  22.6 &  -9.84 &  2.34 & -0.08 &  14.8 & -10.44 &  1.75 & -0.67   \\
  126a &  3.87 &  37.9 &  -5.56 &  5.72 &  1.85 &  50.0 &  -5.56 &  5.72 &  1.85   \\
  126b &  2.82 &  37.2 &  -6.37 &  4.91 &  2.09 &  43.1 &  -6.40 &  4.88 &  2.06   \\
  126c &  3.14 &  31.8 &  -6.26 &  5.02 &  1.88 &  40.5 &  -6.27 &  5.01 &  1.87   \\
  126d &  2.64 &  39.9 &  -6.82 &  4.46 &  1.81 &  44.1 &  -6.88 &  4.40 &  1.76   \\
  126e &  2.96 &  34.3 &  -6.75 &  4.52 &  1.57 &  37.6 &  -6.82 &  4.46 &  1.51   \\
   127$^{*}$ &  1.28 &  20.7 &  -8.50 &  2.10 &  0.83 &  35.6 & -11.42 & -0.82 & -2.09   \\
  128a &  3.37 &  29.7 &  -7.03 &  4.25 &  0.88 &  31.7 &  -7.15 &  4.13 &  0.76   \\
  128b &  3.46 &  34.5 &  -5.74 &  5.54 &  2.09 &  50.0 &  -5.75 &  5.53 &  2.08   \\
  128c &  2.93 &  30.5 &  -7.06 &  4.22 &  1.29 &  31.8 &  -7.11 &  4.17 &  1.23   \\
  128d &  2.83 &  37.4 &  -6.75 &  4.53 &  1.70 &  37.7 &  -6.79 &  4.49 &  1.66   \\
  128e &  2.78 &  33.6 &  -7.11 &  4.17 &  1.39 &  33.4 &  -7.22 &  4.06 &  1.28   \\
  129a$^{*}$ &  1.18 &  29.6 &  -8.82 &  1.85 &  0.68 &  24.1 &  -9.50 &  1.18 &  0.00   \\
  129b$^{*}$ &  1.26 &  31.5 &  -8.70 &  1.97 &  0.72 &  30.4 &  -9.39 &  1.29 &  0.03   \\
  130a &  2.18 &  24.9 &  -8.63 &  2.65 &  0.47 &  28.6 &  -8.76 &  2.52 &  0.33   \\
  130b &  2.21 &  24.9 &  -8.68 &  2.60 &  0.39 &  29.8 &  -8.91 &  2.37 &  0.16   \\
  131a &  3.53 &  39.5 &  -6.84 &  5.24 &  1.71 &  37.0 &  -7.05 &  5.02 &  1.49   \\
  131b &  3.17 &  39.5 &  -6.50 &  5.58 &  2.41 &  38.7 &  -6.73 &  5.34 &  2.17   \\
  131c &  3.32 &  41.4 &  -6.27 &  5.81 &  2.49 &  43.3 &  -6.42 &  5.66 &  2.34   \\
  131d &  3.05 &  41.2 &  -5.98 &  6.10 &  3.04 &  41.3 &  -6.06 &  6.01 &  2.96   \\
  131e &  2.79 &  42.7 &  -6.10 &  5.98 &  3.19 &  43.9 &  -6.14 &  5.93 &  3.14   \\
  131f &  3.38 &  34.0 &  -6.77 &  5.31 &  1.93 &  31.2 &  -6.74 &  5.34 &  1.96   \\
  131g &  3.18 &  31.1 &  -7.12 &  4.95 &  1.77 &  31.9 &  -7.32 &  4.75 &  1.57   \\
  131h &  2.98 &  34.2 &  -7.19 &  4.88 &  1.90 &  32.9 &  -7.44 &  4.63 &  1.65   \\
  131i &  2.45 &  33.5 &  -7.90 &  4.18 &  1.72 &  29.8 &  -8.34 &  3.73 &  1.28   \\
  132a &  3.29 &  40.5 &  -6.46 &  5.61 &  2.32 &  41.3 &  -6.56 &  5.52 &  2.22   \\
  132b &  2.64 &  35.7 &  -6.36 &  5.72 &  3.08 &  42.4 &  -6.44 &  5.64 &  3.00   \\
  132c &  2.27 &  39.1 &  -6.68 &  5.40 &  3.13 &  36.9 &  -6.81 &  5.26 &  2.99   \\
  132d &  3.36 &  39.8 &  -5.86 &  6.22 &  2.86 &  40.6 &  -5.92 &  6.16 &  2.80   \\
  132e &  2.50 &  37.5 &  -6.16 &  5.92 &  3.42 &  30.9 &  -6.21 &  5.87 &  3.37   \\
  134a$^{*}$ &  2.32 &  25.2 &  -9.38 &  1.90 & -0.41 &  20.6 &  -9.89 &  1.39 & -0.92   \\
  134b &  2.03 &  22.7 &  -9.55 &  1.73 & -0.29 &  23.4 &  -9.99 &  1.29 & -0.74   \\
  134c &  1.96 &  22.1 &  -9.43 &  1.84 & -0.11 &  22.1 &  -9.79 &  1.49 & -0.47   \\
  141a &  2.03 &  29.5 &  -8.19 &  3.09 &  1.07 &  31.0 &  -8.36 &  2.92 &  0.89   \\
  141b &  1.63 &  24.5 &  -8.21 &  3.07 &  1.44 &  27.6 &  -8.32 &  2.96 &  1.33   \\
  142a &  2.04 &  35.8 &  -7.56 &  3.72 &  1.68 &  35.9 &  -7.58 &  3.70 &  1.66   \\
  142b &  2.05 &  31.2 &  -7.82 &  3.46 &  1.41 &  32.2 &  -7.90 &  3.38 &  1.33   \\
  144a$^{*}$ &  1.71 &  27.3 &  -9.50 &  1.78 &  0.07 &  23.4 & -10.11 &  1.17 & -0.54   \\
  144b$^{*}$ &  2.04 &  27.7 &  -9.05 &  2.23 &  0.19 &  25.2 & -10.07 &  1.21 & -0.83   \\
  144c$^{*}$ &  1.72 &  24.1 & -10.65 &  0.63 & -1.08 &  35.6 & -11.42 & -0.14 & -1.85   \\
  149a &  2.57 &  33.8 &  -6.73 &  4.55 &  1.98 &  38.4 &  -6.74 &  4.54 &  1.96   \\
  149b &  1.70 &  31.0 &  -7.42 &  3.86 &  2.16 &  31.9 &  -7.50 &  3.78 &  2.08   \\
   150 &  2.45 &  29.9 &  -7.48 &  3.80 &  1.36 &  31.3 &  -7.67 &  3.61 &  1.17   \\
   161 &  2.37 &  34.4 &  -7.39 &  3.89 &  1.52 &  34.2 &  -7.55 &  3.73 &  1.36   \\
   162 &  2.66 &  33.5 &  -7.27 &  4.00 &  1.34 &  33.1 &  -7.42 &  3.86 &  1.20   \\
  163a &  2.78 &  39.4 &  -6.92 &  4.36 &  1.58 &  38.0 &  -6.95 &  4.33 &  1.55   \\
  163b &  2.65 &  39.2 &  -7.27 &  4.01 &  1.36 &  36.2 &  -7.32 &  3.96 &  1.31   \\
  163c &  2.76 &  35.8 &  -7.13 &  4.15 &  1.39 &  36.8 &  -7.30 &  3.98 &  1.21   \\
  165a &  2.53 &  35.1 &  -7.44 &  3.84 &  1.31 &  34.6 &  -7.54 &  3.74 &  1.21   \\
  165b &  2.68 &  33.3 &  -6.96 &  4.31 &  1.63 &  34.8 &  -7.04 &  4.24 &  1.55   \\
  167a &  2.26 &  30.3 &  -7.61 &  3.67 &  1.41 &  29.3 &  -7.70 &  3.58 &  1.31   \\
  167b &  2.64 &  32.5 &  -7.08 &  4.20 &  1.55 &  32.8 &  -7.19 &  4.09 &  1.45   \\
  167c &  2.17 &  31.1 &  -7.56 &  3.72 &  1.54 &  30.6 &  -7.70 &  3.58 &  1.41   \\
   183 &  2.78 &  23.6 &  -8.95 &  2.91 &  0.13 &  31.2 &  -9.37 &  2.49 & -0.29   \\
   185 &  2.08 &  46.3 &  -7.45 &  4.41 &  2.33 &  41.8 &  -7.46 &  4.40 &  2.33   \\
   188$^{*}$ &  2.72 &  23.2 &  -8.86 &  3.00 &  0.28 &  17.3 &  -9.25 &  2.62 & -0.11   \\
  190a &  2.75 &  27.1 &  -8.24 &  3.62 &  0.87 &  27.8 &  -8.31 &  3.55 &  0.80   \\
  190b &  2.74 &  27.2 &  -8.02 &  3.84 &  1.10 &  30.1 &  -8.05 &  3.81 &  1.07   \\
  199a &  2.89 &  26.3 &  -8.05 &  3.81 &  0.92 &  29.6 &  -8.19 &  3.68 &  0.79   \\
  199b &  2.26 &  24.2 &  -7.65 &  4.22 &  1.95 &  19.1 &  -8.19 &  3.68 &  1.41   \\
  201a &  2.71 &  29.9 &  -7.95 &  3.91 &  1.20 &  28.7 &  -8.22 &  3.65 &  0.94   \\
  201b &  2.91 &  29.8 &  -7.51 &  4.35 &  1.44 &  29.2 &  -7.77 &  4.09 &  1.18   \\
  202a &  3.52 &  24.0 &  -8.89 &  2.98 & -0.55 &  24.7 &  -9.29 &  2.58 & -0.95   \\
  202b$^{*}$ &  2.79 &  24.8 &  -8.84 &  3.02 &  0.23 &  44.4 & -10.20 &  1.66 & -1.13   \\
  202c &  3.02 &  31.6 &  -7.60 &  4.27 &  1.25 &  31.6 &  -7.65 &  4.22 &  1.20   \\
  202d &  2.88 &  25.7 &  -8.23 &  3.63 &  0.75 &  25.1 &  -8.42 &  3.44 &  0.56   \\
  202e &  2.67 &  25.9 &  -8.58 &  3.28 &  0.62 &  27.3 &  -8.76 &  3.10 &  0.43   \\
  202f$^{*}$ &  2.71 &  25.3 &  -9.12 &  2.74 &  0.03 &  35.6 & -11.42 &  0.45 & -2.26   \\
  202g$^{*}$ &  2.73 &  26.1 &  -8.80 &  3.07 &  0.34 &  24.0 &  -9.56 &  2.31 & -0.42   \\
  202h$^{*}$ &  3.00 &  23.3 & -10.32 &  1.54 & -1.46 &  10.0 & -10.32 &  1.54 & -1.46   \\
  202i$^{*}$ &  2.92 &  23.0 &  -9.25 &  2.61 & -0.31 &  35.6 & -11.42 &  0.45 & -2.48   \\
  203a &  2.92 &  32.7 &  -7.89 &  3.98 &  1.06 &  35.9 &  -7.92 &  3.94 &  1.02   \\
  203b &  2.79 &  25.1 &  -8.32 &  3.55 &  0.76 &  24.8 &  -8.40 &  3.46 &  0.67   \\
  203c &  3.21 &  31.6 &  -7.56 &  4.30 &  1.09 &  31.7 &  -7.58 &  4.28 &  1.07   \\
  203d &  2.90 &  34.5 &  -7.65 &  4.21 &  1.31 &  35.3 &  -7.67 &  4.20 &  1.29   \\
  208a &  3.18 &  23.8 &  -8.17 &  3.69 &  0.51 &  20.8 &  -8.27 &  3.59 &  0.41   \\
  208b &  2.51 &  23.3 &  -8.90 &  2.96 &  0.45 &  23.8 &  -9.09 &  2.77 &  0.26   \\
\enddata
\tablenotetext{a}{Here $\rm T_{\rm c, tot}$ and $L_{\rm tot}$ are the colder component 
temperature and infrared luminosity using the non-background subtracted method. 
While $\rm T_{\rm c}$ and $L$ are derived using the background subtracted 
method.}
\tablenotetext{b}{The mass $\log(M)$ is from Paper I.}
\tablenotetext{*}{Clumps with uncertain measurements
of $L$ due to IRAS 100\micron\ background subtraction (see Fig.~\ref{Ldist}b).}
\end{deluxetable}

\begin{deluxetable}{ccc cc cc cc cc cc}
\rotate
\tabletypesize{\footnotesize}
\tablecolumns{13}
\tablewidth{0pc}
\tablecaption{Secondary physical properties of the HCO$^+$(1-0) clumps.$^a$ \label{tab:clumps2}}
\tablehead{
\colhead{BYF} & 
\colhead{ $\log(F_{\rm w})$  }   & \colhead{$\log(F_{\rm w, tot})$} & 
\colhead{ $\log(F_{\rm IRAC})$  }   &\colhead{ $\log(F_{\rm IRAC, tot})$  } & 
\colhead{ $\log(I_{\rm IRAC})$  }   &\colhead{ $\log(I_{\rm IRAC, tot})$  } & 
\colhead{ $T_{\rm bol}$  }   & \colhead{ $T_{\rm bol, tot}$  }   & 
\colhead{ $\Sigma$  }   & \colhead{ $\sigma_{\rm v}$  }   & 
\colhead{ $\alpha_{\rm vir}$  }  & \colhead{ $\log(L_{\rm HCO^+(1-0)})$  } 
\\ 
\colhead{No.} & 
\colhead{$\rm erg/s/cm^{2}$ }  &\colhead{$\rm erg/s/cm^{2}$ }  & 
\colhead{$\rm erg/s/cm^{2}$ }  &\colhead{$\rm erg/s/cm^{2}$ } & 
\colhead{$\rm erg/s/cm^{2}/sr$ } &\colhead{$\rm erg/s/cm^{2}/sr$ }  & 
\colhead{ (K) }   & \colhead{ (K) }   & 
\colhead{ $\rm g/cm^2$  }   & \colhead{ $\rm km/s$  }   & 
\colhead{ $$  }  & \colhead{$\rm K \; km/s \; pc^{2}$ }
}


\startdata
    5a &  -7.65 &  -7.68 &  -8.68 &  -8.50 &  -2.78 & -2.60 &  99 &  93 & 0.033 &  1.90 &   6.1 &  1.43 \\
    5b &  -8.94 &  -8.80 &  -9.95 &  -9.47 &  -3.23 & -2.75 & 106 & 104 & 0.029 &  3.10 &  54.4 &  0.37 \\
    5c &  -8.57 &  -8.59 &  -9.44 &  -9.18 &  -3.21 & -2.95 & 107 & 108 & 0.029 &  3.00 &  26.5 &  0.76 \\
    5d &  -7.42 &  -7.48 &  -7.84 &  -7.81 &  -2.06 & -2.03 & 138 & 136 & 0.022 &  2.60 &  14.7 &  1.15 \\
\enddata
\tablenotetext{a}{We only show the first 4 entries to show the format of this table. $\sigma_{\rm v}$ and $\alpha_{\rm vir}^{b}$ are from Paper I. The whole table 
will be available online only.}

\end{deluxetable}


\end{document}